\documentclass[aps,prb,preprint]{revtex4-1}

\setcitestyle{super,open={},close={}}

\usepackage[utf8]{inputenc}
\usepackage[T1]{fontenc}
\usepackage[english]{babel}
\usepackage[version=4]{mhchem}
\usepackage{siunitx}
\usepackage{graphicx}

\usepackage{geometry}
\usepackage{caption}
\usepackage{natbib}
\usepackage{setspace}
\usepackage{xkeyval}
\usepackage{array}
\usepackage{booktabs}
\usepackage{listings}
\usepackage{lmodern}
\usepackage{mathpazo}
\usepackage{microtype}
\usepackage{hyperref}

\usepackage{amsmath}
\usepackage{amssymb}
\usepackage{amsfonts}
\usepackage{latexsym}
\usepackage[compact]{titlesec}
\usepackage[usenames,dvipsnames]{xcolor}
\usepackage{mathptmx}
\usepackage{calc}
\usepackage{titletoc}
\usepackage{lipsum}
\usepackage{multirow}
\usepackage{amstext}
\usepackage{tabularx}
\usepackage{adjustbox}
\usepackage{epsfig,color}
\usepackage{placeins}
\usepackage{natmove}
\usepackage{titlesec}
\usepackage{tablefootnote}
\usepackage{threeparttable}
\usepackage[d]{esvect}
\usepackage{mathtools}
\usepackage{textcomp}
\usepackage{floatrow}
\usepackage{subcaption}

\usepackage[none]{hyphenat}
\usepackage{microtype}
\usepackage{xcolor}

\vfuzz \hfuzz
\begin{document}
\title{Signature of excitonic insulators in phosphorene nanoribbons}

\author{A. F. P. de Oliveira}
\author{Andr\'eia L. da Rosa}
\email{andreialuisa@ufg.br}
\affiliation{Institute of Physica, Federal University of Goi\'as, Campus Samambaia, 74690-900, Goi\^ania, Goi\'as, Brazil.}

\author{Alexandre C. Dias}
\email{alexandre.dias@unb.br}
\affiliation{Institute of Physics and International Center of Physics, University of Bras{\'i}lia, Bras{\'i}lia $70919$-$970$, Distrito Federal, Brazil}

\begin{abstract}

Phosphorene is a recently developed two-dimensional (2D) material that has attracted tremendous attention because of its unique anisotropic optical properties and quasi-one-dimensional  (1D) excitons. We use first-principles calculations combined with the maximally localized Wannier function tight binding  Hamiltonian (MLWF-TB) and Bethe-Salpeter equation (BSE) formalism to investigate quasiparticle effects of 2D and quasi-1D blue and black phosphorene nanoribbons.  Our electronic structure calculations shows that both blue and black monolayered phases are semiconductors. On the other hand black phosphorene zigzag nanoribbons are metallic. Similar behavior is found for very thin blue phosphorene zig-zag and armchair nanoribbon. As a general behavior, the exciton binding energy decreases as the ribbon width increases, which highlights the importance of quantum confinement effects. The solution of the BSE shows that the blue phosphorene monolayer has an exciton binding energy four times higher than that of the black phosphorene counterpart. Furthermore, both monolayers show a different linear optical response with respect to light polarization, as black phosphorene is highly anisotropic. We find a similar, but less pronounced, optical anisotropy for blue phosphorene monolayer, caused exclusively by the quasi-particle effects. Finally, we show that some of the investigated nanoribbons show a spin-triplet excitonic insulator behavior, thus revealing exciting features of these nanoribbons and therefore provides important advances in the understanding of quasi-one dimensional phosphorus-based materials.

\end{abstract}

\maketitle

\section{Introduction}

Few-layer black phosphorus has recently attracted attention because of its fascinating applications in several technological fields, including electronic and optoelectronic devices\,\cite{PhysRevB.89.235319,Chaudhary_2022,Tareen2022} Monolayer phosphorene has been of particular interest in exploring technological applications and investigating fundamental phenomena, such as 2D quantum confinement and many-body interactions. However, phosphorene monolayers are usually unstable against oxidation and environmental attack under ambient conditions, making its characterization challenging.\cite{NJP2D2021} Consequently, many important physical properties of phosphorene, such as its excitonic nature, remain unclear.  

Black phosphorus, an elemental phosphorus allotrope, was
first synthesized in $1914$ and was recently rediscovered as an
exciting member of the 2D family,\cite{6}. In black phosphorene, each atom is bonded to three adjacent phosphorus atoms in a puckered structure. It is possible to synthesize a monolayer black phosphorene via chemical exfoliation, since bulk black phosphor is composed of single atomic layers held together by van der Waals interactions. Black phosphorus has a direct bandgap, which can be tuned from \SI{0.3}{\electronvolt} (bulk) to \SI{2.0}{\electronvolt} (monolayer) as a result of layer-dependent photoluminescence (PL) spectra.\cite{Yang2015}  Single layer black phosphorene has quasi-particle and optical gaps of \SI{2.2}{\electronvolt} measured by photoluminescence spectroscopy.\cite{NNano15} Near-band-edge recombinations are observed at \SI{2}{\kelvin}, including dominant excitonic transitions at \SI{0.276}{\electronvolt} and a weaker one at \SI{0.278}{\electronvolt}.\cite{Carre_2021} This makes black phosphorene ideal for optoelectronic applications such as broadband photodetection.\cite{18} First-principles GW-BSE calculations shows the effect of strain on excitons in pristine black phospherene monolayers\cite{PhysRevB.99.045432} reported \SI{1.40}{\electronvolt} for the optical bandgap and \SI{0.74}{\electronvolt} for the exciton binding energy. Experimental results suggested \SI{0.9}{} and \SI{0.12}{\electronvolt}.\cite{Wang2018}
More recently a linearly polarized luminescence emission owing to from edge reconstructions has been identified\,\cite{Souvik2023} However, some of the structures shown in \,\cite{Souvik2023} reconstruct whereas other do not and therefore the condition for  edge reconstruction has no been completely clarified. Here we do not find any spontaneous reconstruction of the edges; thus, strain plays a minor role.

In addition to the black structure, phosphorene can be crystallized in hexagonal symmetry (blue phosphorene). This structure is the most stable form on noble metal substrates such as \ce{Au}($111$) and \ce{Ag}($111$), while black phosphorene is the most stable
phase on metal surfaces such as \ce{Al}($111$) and \ce{Sn}($100$)\,\cite{47}. Similarly to black phosphorene, the fundamental bandgap of blue phosphorene
depends on the number of layers. Recently, the experimental electronic bandgap structure of blue phosphorene on \ce{Au}($111$) substrate was determined to be \SI{1.1}{\electronvolt}\,\cite{Zhuang2018}.

In addition to two-dimensional nanostructures, more recently quasi-one-dimensional phosphorene has been reported. These long phosphorene chains were synthesized on an Ag (111) surface by molecular beam epitaxy (MBE) and characterized by scanning tunneling microscopy, showing a band gap of \SI{1.5}{\electronvolt}, which makes them viable for possible applications in nanoelectronics.\cite{zhang2021flat}.   In low-dimensional structures, the screening of the Coulomb interaction is strongly reduced, which can have important consequences such as the significant increase of exciton binding energies. Since reduced electron screening in two plays a fundamental role in determining exciton properties, understanding the optoelectronic and photonic properties of phosphorene is fundamental in determining its device performance. 

In this work we use density functional theory combined with maximally localized wannier functions tight binding (MLWF-TB) Hamiltonian and Bethe-Salpeter formalism to investigate the  the excitonic properties of pristine phosphorene nanosheets and  nanoribbons. Our calculations revealed strongly bound excitons with anisotropic spatial delocalization. We show that the exciton binding energies in both sheets and ribbons are highly anisotropic.  Furthermore, we suggest that phosphorene nanoribbons are an excitonic insulator, arising from the spontaneous formation of electron-hole-bound states.  Our results provide a new feature of these materials and help to understand the ground-state excitons. Localized emission from black phosphorous edges motivates exploration of nanoribbons  for tunable narrow band light generation, with future potential to create nanostructures for quantum information processing applications as well as exploration of exotic phases.

\section{Methodology}

First-principle spin-polarized calculations have been performed using density functional theory (DFT), within the generalized gradient approximation (GGA) according to the Perdew-Burke-Ernzerhof (PBE) parameterization,\cite{perdew1996generalized} to determine the structural properties and phonon dispersion of materials based on phosphorus. As DFT-PBE tends to underestimate the fundamental band gap due to self-interaction errors,\cite{Cohen_115123_2008,Crowley_1198_2016} we have used  range-separated hybrid exchange-correlation functional proposed by Heyd, Scuseria and Ernzerhof (HSE06)\cite{heyd2004efficient,hummer2009heyd} to improve the electronic properties of the investigated nanostructures.

The projector augmented wave (PAW) method,\cite{kresse1999ultrasoft} as implemented in the Vienna\,\textit{ab initio} simulation package (VASP)\,\cite{Kresse_13115_1993,Kresse_11169_1996} has been used to describe the electronic charge density. To obtain the electronic and structural properties, we chose a plane wave energy cutoff of \SI{400}{\electronvolt}, with a total energy convergence criterion of \SI{E-6}{\electronvolt}. The ground state properties have been obtained until the atomic forces were less than \SI{0.01}{\electronvolt \per \angstrom}. These calculations were done considering a \textbf{k}-points density of \SI{40}{\per \angstrom} in each periodic lattice vector direction. Further details on the atomic structures and the vacuum in the non-periodic directions can be seen in the Support Information (SI).

Phonon dispersion has been obtained combining calculations performed with VASP and Phonopy package,\cite{togo2008first}. We chose the density functional perturbation theory (DFPT) method, with a $4\times3\times1$($4\times4\times1$) supercell for the pristine black(blue) phosphorene monolayer. We use the same energy cutoff and automatic Monkhorst-Pack \cite{monkhorst1976special} \textbf{k}-mesh generation scheme of the electronic properties.

In order to investigate the excitonic and optical properties, we first generate a maximally localized Wannier function tight-binding (MLWF-TB) Hamiltonian, directly obtained from DFT-HSE06 calculations using the Wannier90 package.\cite{mostofi2008wannier90} Then we investigate the linear optical response at independent particle approximation (IPA). Next, to include electron-hole interaction we solve the Bethe-Salpeter equation (BSE)\,\cite{Salpeter_1232_1951,Dias_085406_2020}  using the Rytova-Keldysh 2D potential (V2DRK).\cite{VanTuan_125308_2018} The nanostructures are surrounded by a vacuum region. We use a \SI{120}{\per\angstrom} \textbf{k}-points density and a smearing of \SI{0.05}{\electronvolt} for the calculation of real and imaginary parts of the frequency dependent dielectric tensor. Further details about the parameters used in the BSE calculations are given in the SI. We define the optical band gap as the lowest optical transition, with an excitation energy greater than or equal to \SI{0}{\electronvolt}, with an optical transition rate greater than \SI{0.1}{\square\angstrom}. Both IPA and BSE linear optical response levels are obtained using WanTiBEXOS implementation.\cite{Dias_108636_2022}

\section{Results}

\subsection{Structural and Electronic Properties}

\subsubsection{Pristine phosphorene monolayers}

\begin{figure}[H]
    \centering
  \includegraphics[clip,width=0.9\linewidth]{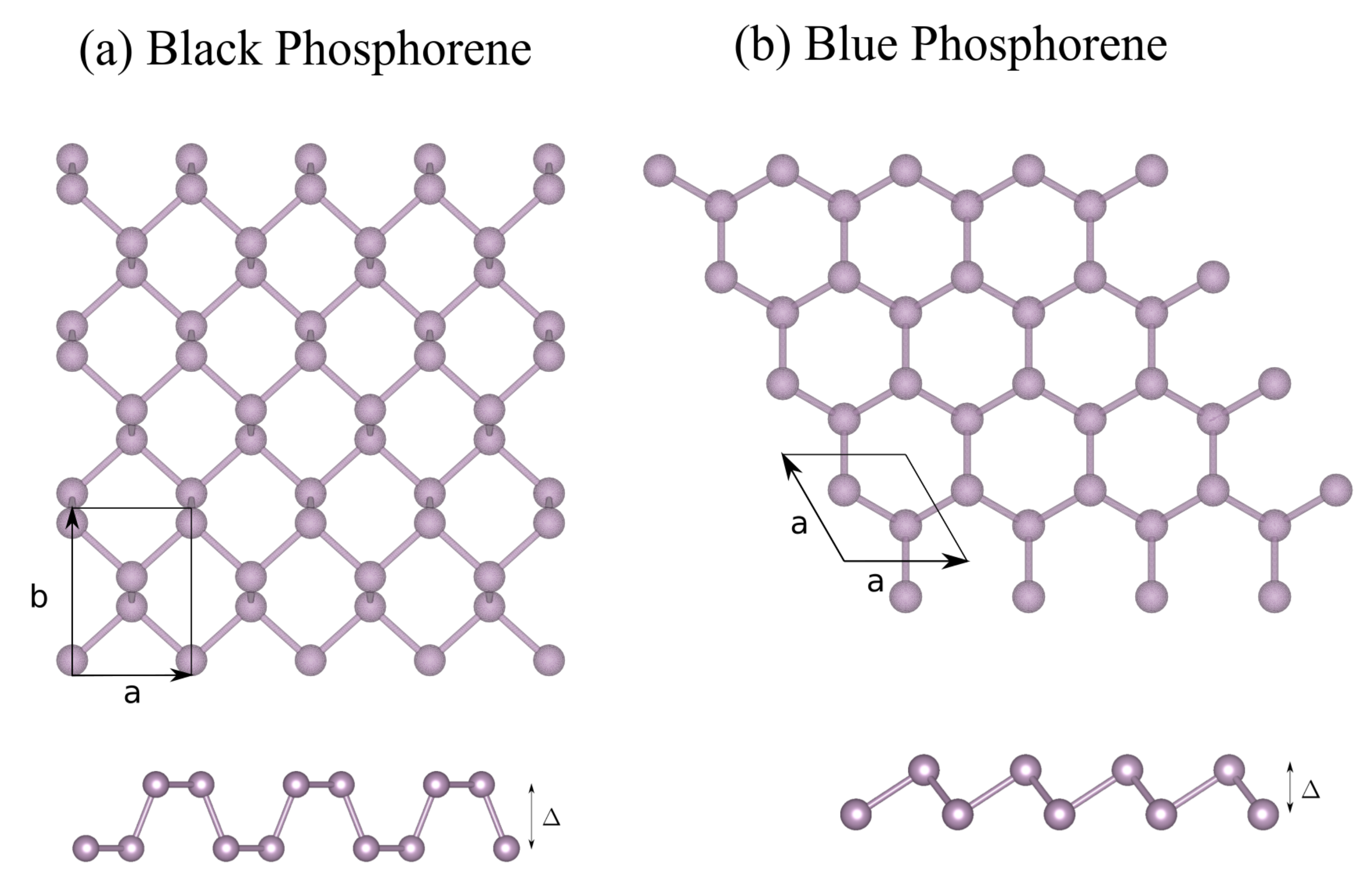}
    \caption{Top (upper panel) and side view (lower panel) of (a) black and (b) blue phosphorene monolayer. Buckling $\Delta$ and unit cell lattice parameters $a$ and $b$ are shown.}\label{fig:crystal_mono}
\end{figure}

As a matter of completeness we show the band structure for 
Black phosphorene monolayer has a puckered structure as shown in Fig.\,\ref{fig:crystal_mono}(a). The relaxed lattice parameters are $ a =4.62$ \si{\angstrom} and $ b = 3.30$ \si{\angstrom}.  Free-standing black phosphorene is not planar, but shows a corrugation $ \Delta = 2.10$ \si{\angstrom} as seen in Fig.\,\ref{fig:crystal_mono}(a)(lower panel). Our results are in agreement with previous works.\cite{Souvik2023,Kaddar2023,Carre_2021}

\begin{figure}[H]
    \centering
  \includegraphics[width=0.9\linewidth]{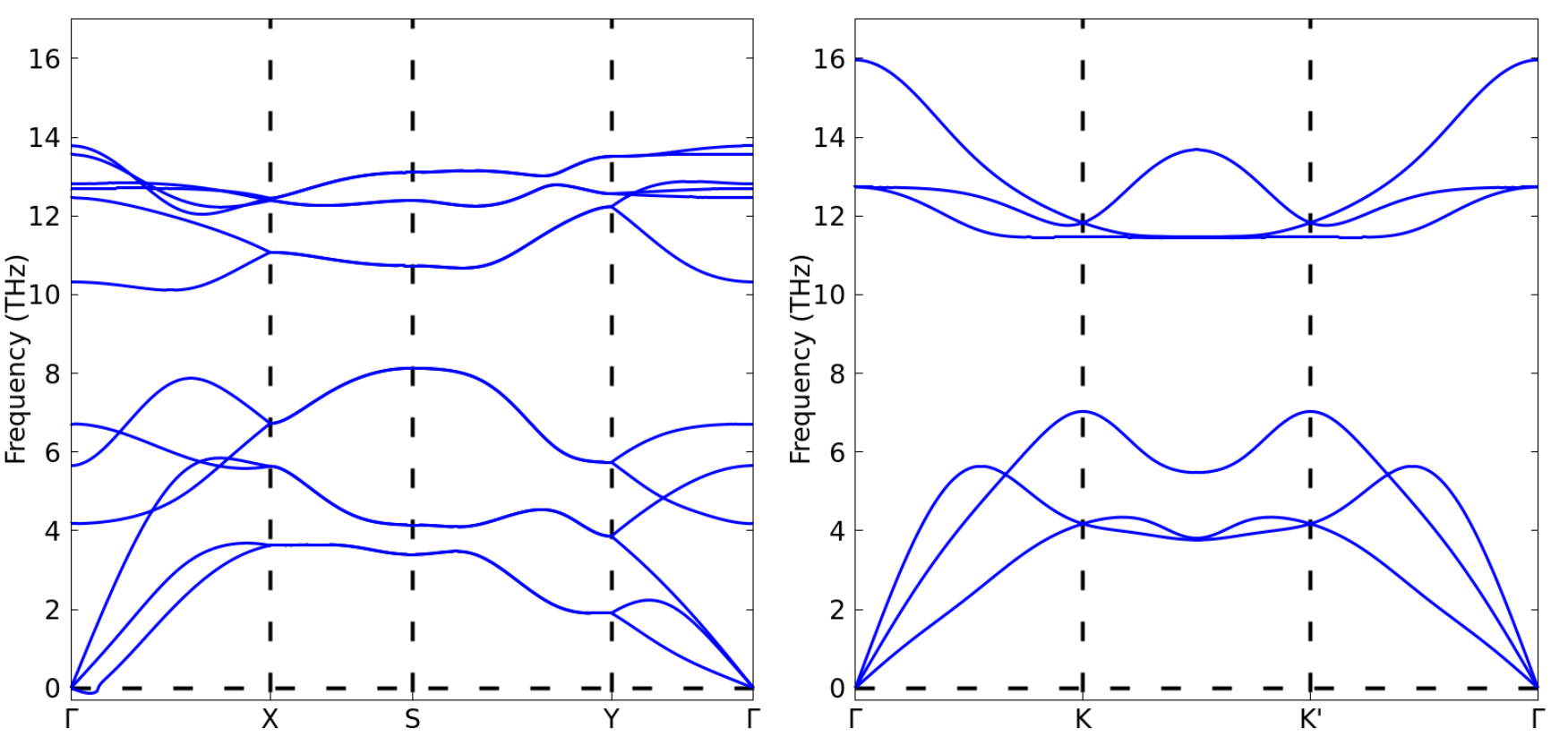}\\
  \includegraphics[width=0.9\linewidth]{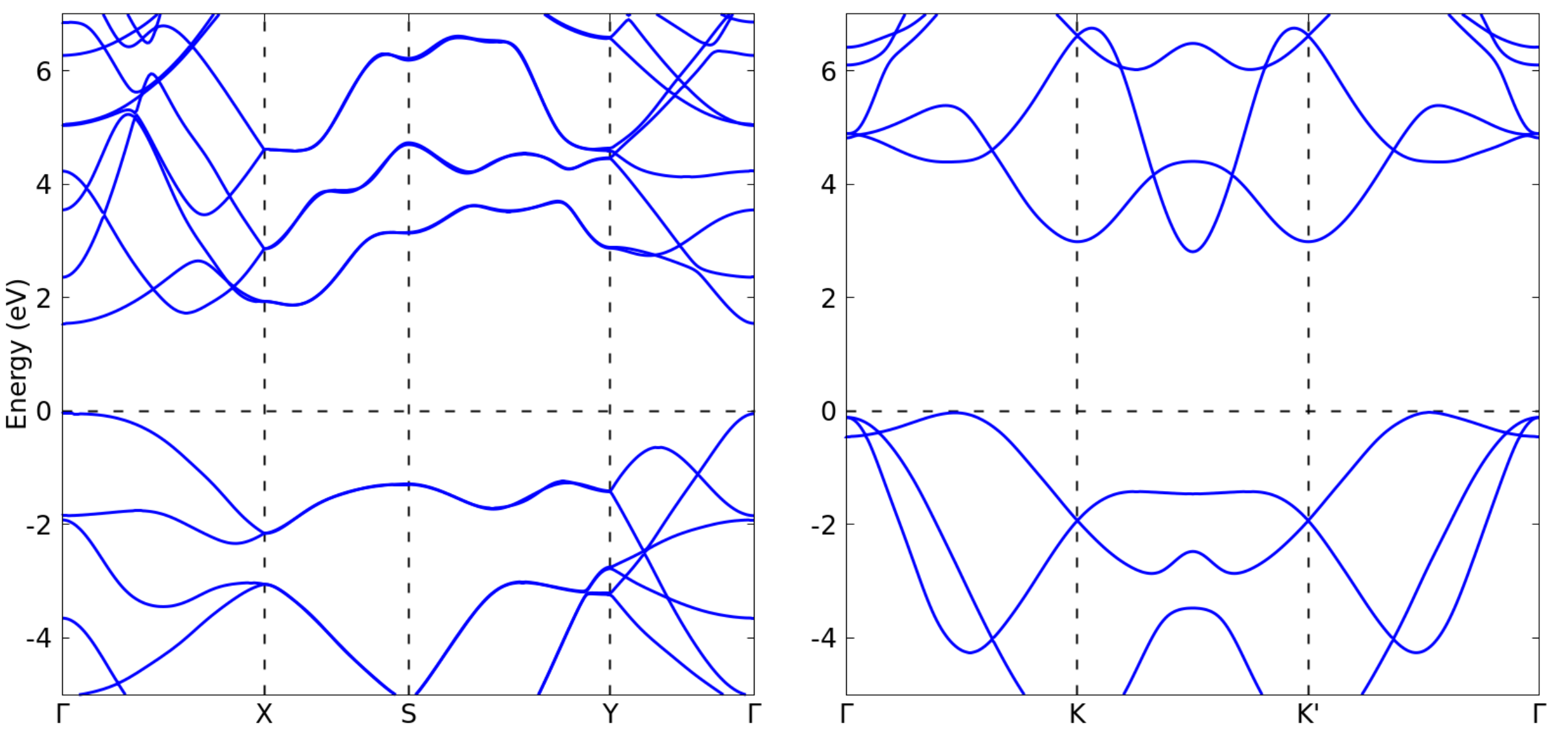}
    \caption{Electronic  band structure and phonon dispersion calculated within HSE06 for (a) black and (b) blue phosphorene. The Fermi level is set at zero.}\label{fig:bands_mono}
\end{figure}

On the other hand, blue phosphorene has a honeycomb structure shown
in Fig.\,\ref{fig:crystal_mono}(b). The relaxed lattice parameters are a= b= \SI{3.28}{\angstrom}. The buckling is $\Delta$ = \SI{1.23}{\angstrom} (lower panel).

Black phosphorene has been found to have an anisotropic behavior in its thermal and electrical properties as reported previously\,\cite{jain2015strongly}. This anisotropic behavior can be verified on phonon dispersion relations of black phosphorene as seen in Fig.\,\ref{fig:bands_mono}(a), right panel. The phononic band gap is \SI{2.5}{\tera\hertz}. The slope of the acoustic branches along the Y-$\Gamma$ (armchair) and $\Gamma$-X (zig-zag) directions highlights the thermodynamic anisotropy of this structure.

Once the atomic positions and lattice vectors  have been optimized, we calculated the electronic band structure as shown in Fig.\,\ref{fig:bands_mono}(lower left panel). We obtained a band gap of \SI{0.91}{\electronvolt} with PBE and \SI{1.60}{\electronvolt} with HSE06 functional.

Blue phosphorene also shows an isotropic behavior for thermal and
electrical properties, measured by in Ref.\,\cite{jain2015strongly}. This isotropic behavior can be verified on phonon dispersion relations of blue phosphorene shown in Fig.\,\ref{fig:bands_mono} (lower right panel). The phonon band gap is \SI{4.9}{\tera\hertz} with the slope of the acoustic branches being anisotropic. The computed electronic band gap is \SI{1.93}{\electronvolt} with PBE and \SI{2.83}{\electronvolt} with HSE06. HSE06 calculations shows a better agreement with previous work\,\cite{Souvik2023,Kaddar2023,Carre_2021,NNano15} and will be therefore used in the calculations of properties of nanoribbons as shown next.

\subsection{Phosphorene  nanoribbons}

\begin{figure}[ht!]
\centering
\includegraphics[width=1\textwidth,clip]{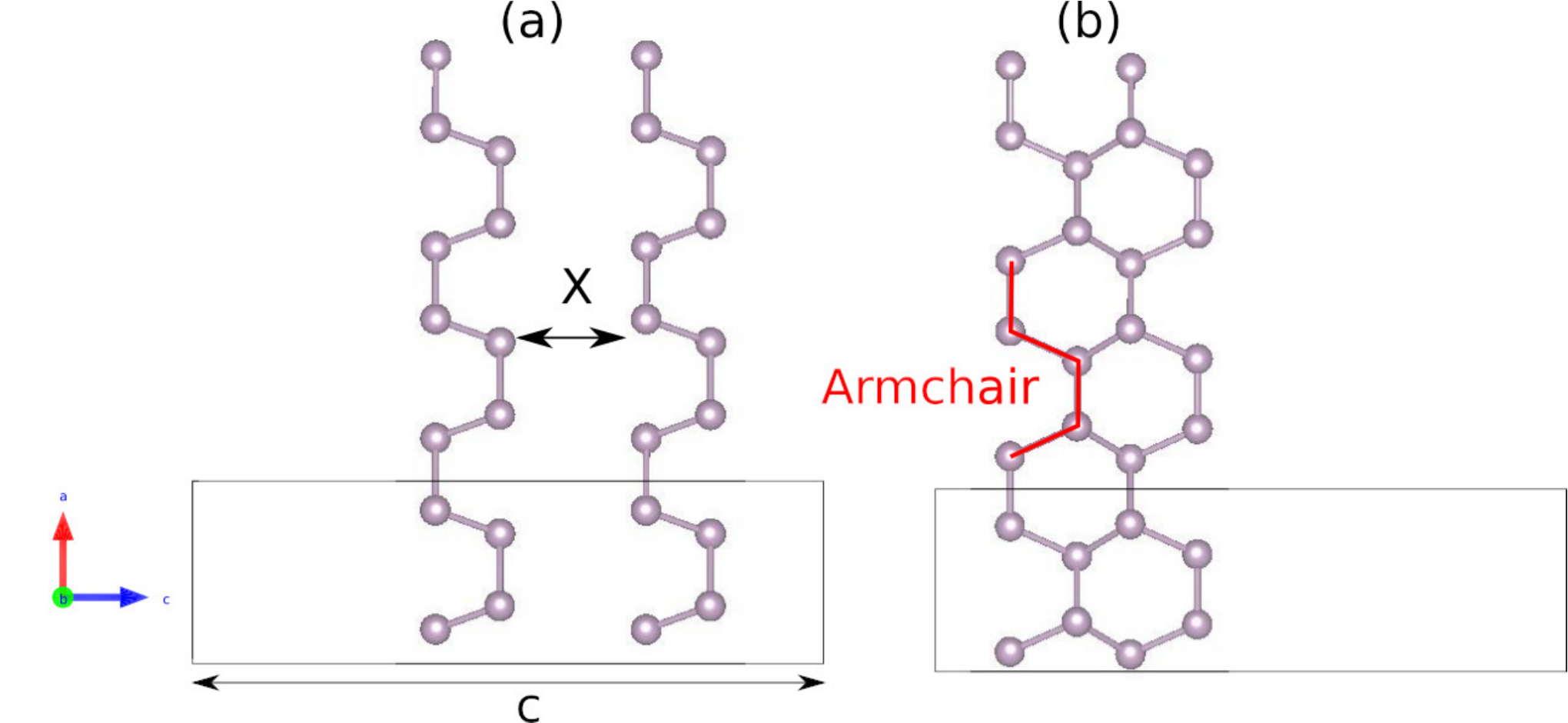}
\caption{\label{fig:twochains}(a) Schematic view of ribbon assembling. Top view before and (b) after atomic optimization of two interacting nanochains for distance X less than \SI{4.0}{\angstrom}. Here the ribbon has armchair edges.}
\end{figure}

Quasi-one-dimensional phosphorene nanostructures have successfully been fabricated on \ce{Ag}($111$) substrates\,\cite{zhang2021flat}. Those structures are composed
of armchair chains made of black phosphorene as seen in Fig.\,\ref{fig:twochains}. 
Placing two nanochains of phosphorus oriented along the zigzag direction as shown in Fig.\,\ref{fig:twochains}(a) can lead to spontaneous formation of nanoribbons. We find that up to a distance of \SI{4.00}{\angstrom} the nanochains relax in such a way as to form ribbons as shown in Fig.,\ref{fig:twochains}(b). This ribbon resembles blue pristine phosphorene with armchair orientation.  We notice that we have investigated several possible configurations, and the ones shown here have the lowest energy among them.

\begin{figure}[ht!]
\centering
\includegraphics[width=0.8\textwidth,clip]{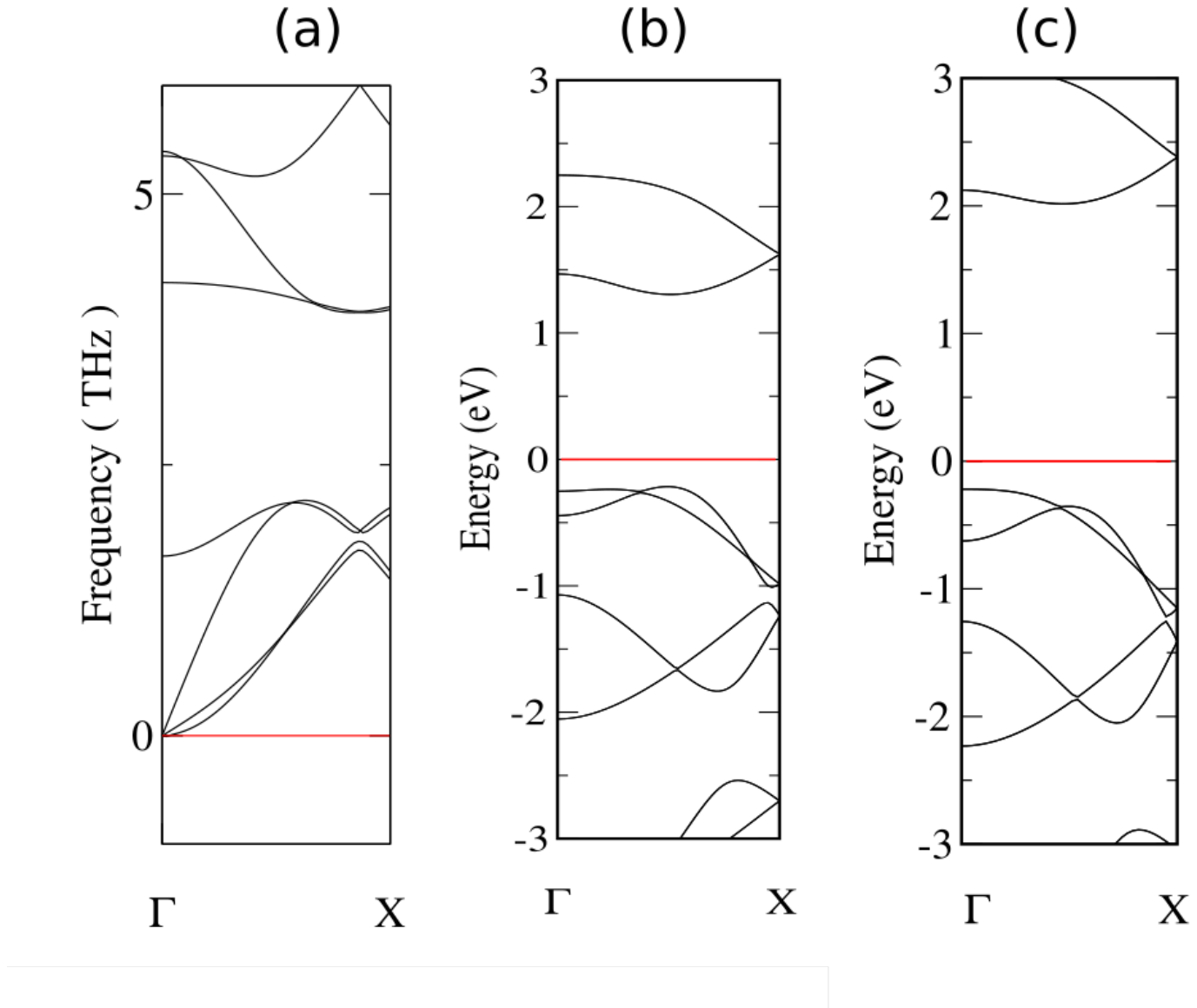}

\caption{\label{fig:chain_phonon_band}(a) Phonon dispersion of a thin phosphorene ribbon shown in Fig.\,\ref{fig:twochains}. Band structure calculated with b) PBE and c) HSE06 exchange-correlation functionals.}
\end{figure}

Phonon dispersion allows us to verify that for this geometry no imaginary frequencies are found in Fig.\,\ref{fig:chain_phonon_band} (a), which indicates the thermodynamic stability of the structure at low temperatures. Acoustic and optical branches are seen, and mixed modes are also seen. The band structure and phonon dispersion are shown in Fig.\,\ref{fig:chain_phonon_band}. The evaluated band gap is \SI{2.20}{\electronvolt} with HSE06 as seen in Figs.\,\ref{fig:chain_phonon_band}(b).  Our calculations show that HSE06 overestimates the experimental gap of \SI{1.8}{\electronvolt}.\cite{zhang2021flat} A possible explanation for this discrepancy is the effect of the substrate reported experimentally, whose effect is out of the scope of this paper.

When discussing nanoribbon electronic and optical properties, it is important to investigate exciton quantum confinement effects. Therefore the ribbons width L was varied in order verify strain and reconstruction effects. In general, we find that strain plays a minor role in these nanostructures, and spontaneous reconstruction does not occur.

\begin{figure}[H]
    \centering
    \subfloat[a)]{\includegraphics[width=0.9\linewidth]{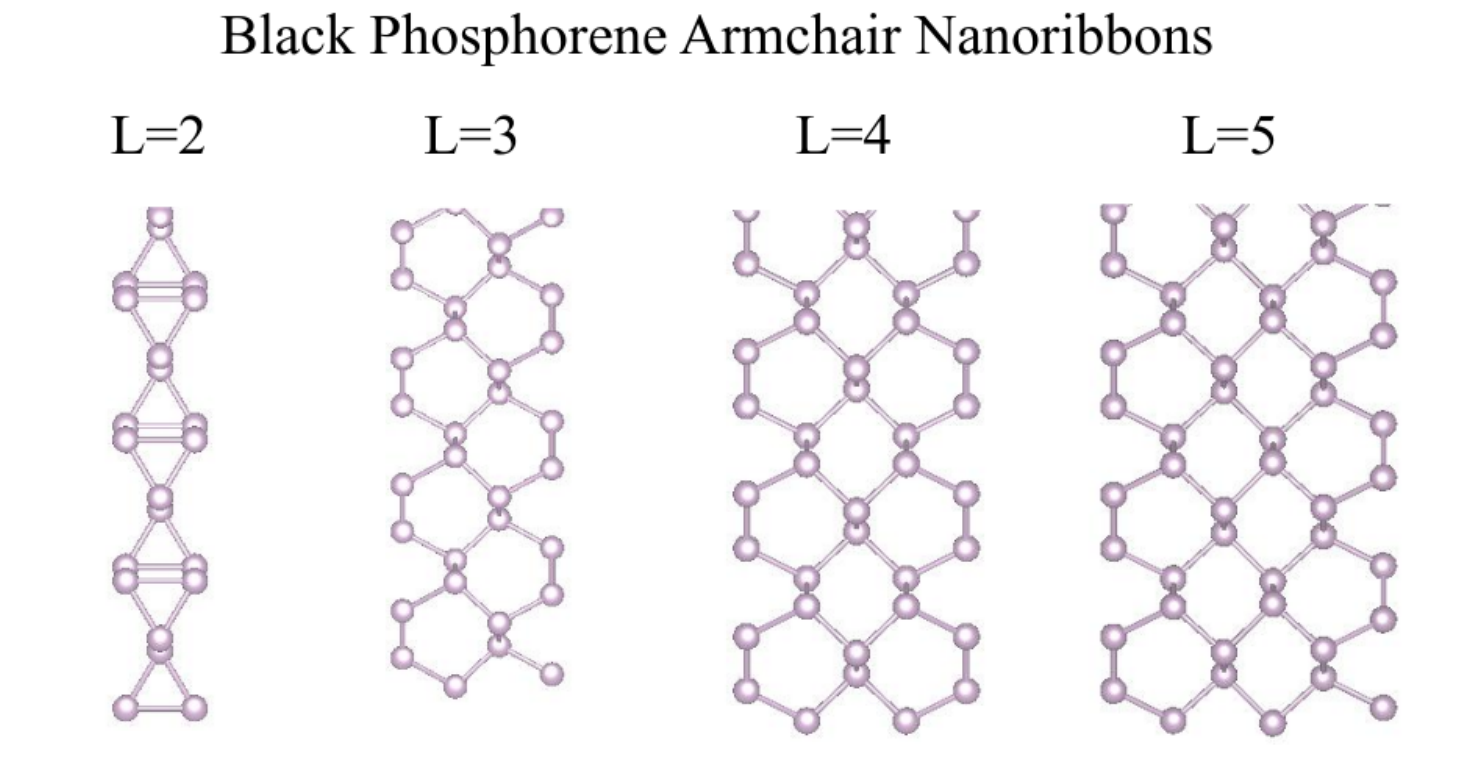}}\\
     \subfloat[b)]{\includegraphics[width=0.9\linewidth]{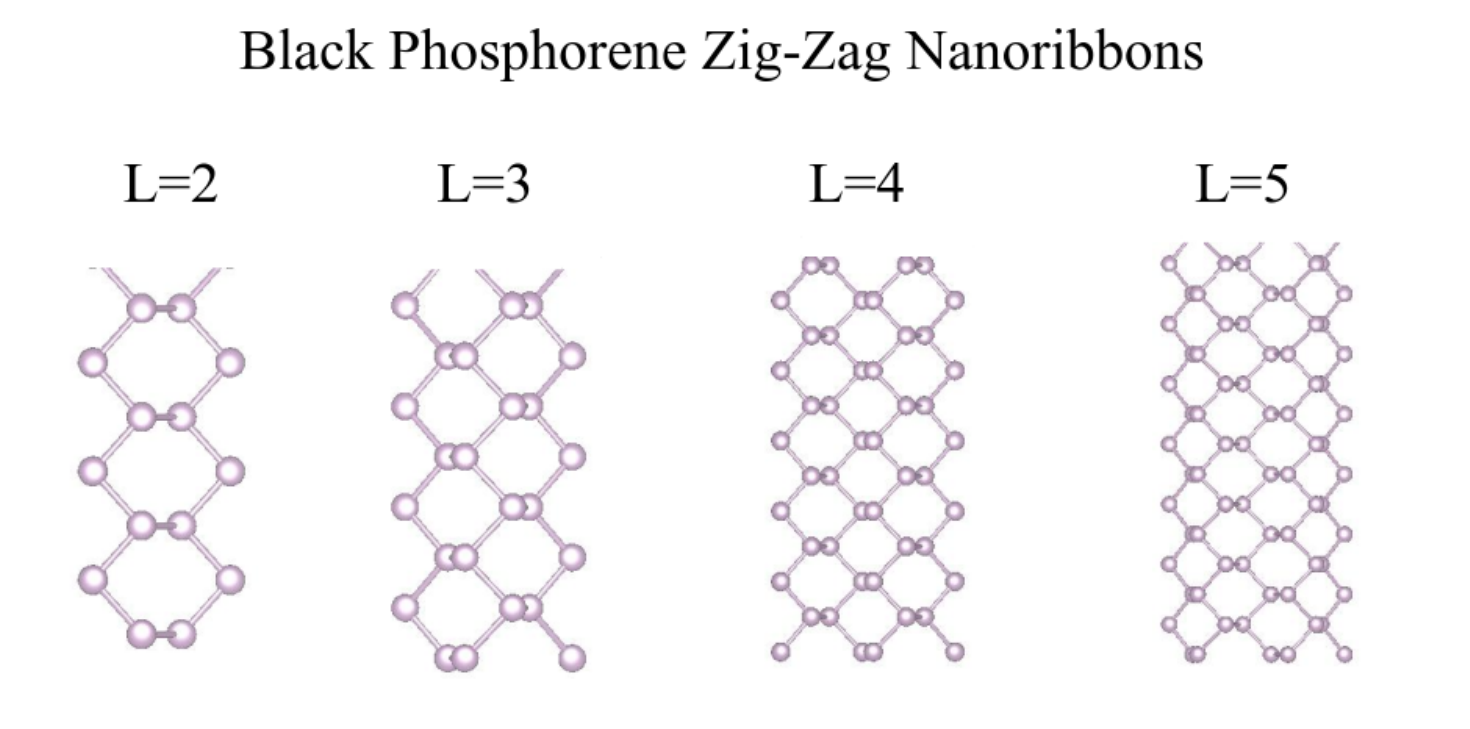}}
    \caption{Relaxed geometries of black phosphorene nanoribbons. a) armchair termination and widths of \SI{2.5}{(L=2)}, \SI{5.2}{(L=3)}, \SI{6.7}{(L=4)} and \SI{8.2}(L=5){\AA}. b) zigzag termination and widths  of \SI{4.8}{(L=2)}, \SI{7.0}{(L=3)}, \SI{8.5}{(L=4)} and \SI{12.5}(L=5){\AA}.}
    \label{fig:black_ribbons}
\end{figure}

\begin{figure}[H]
    \centering
     \includegraphics[width=0.9\linewidth]{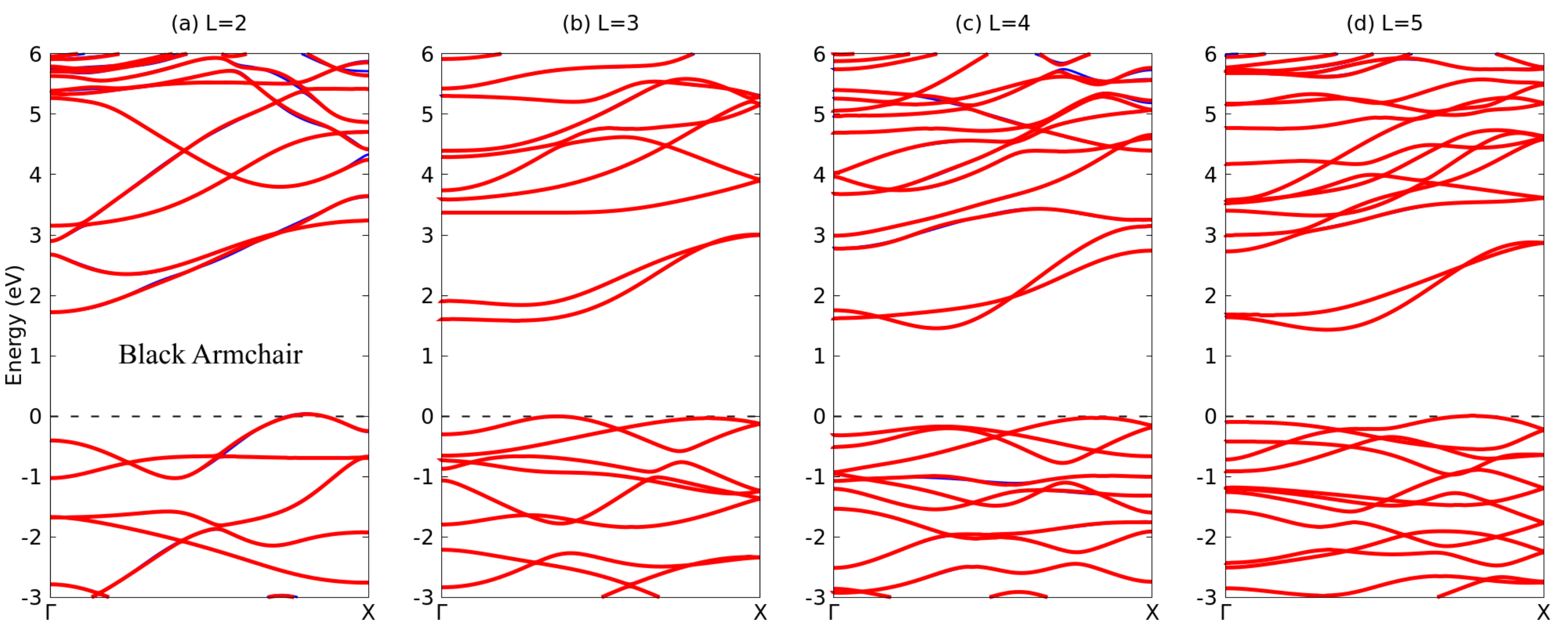}\\
    \includegraphics[width=0.9\linewidth]{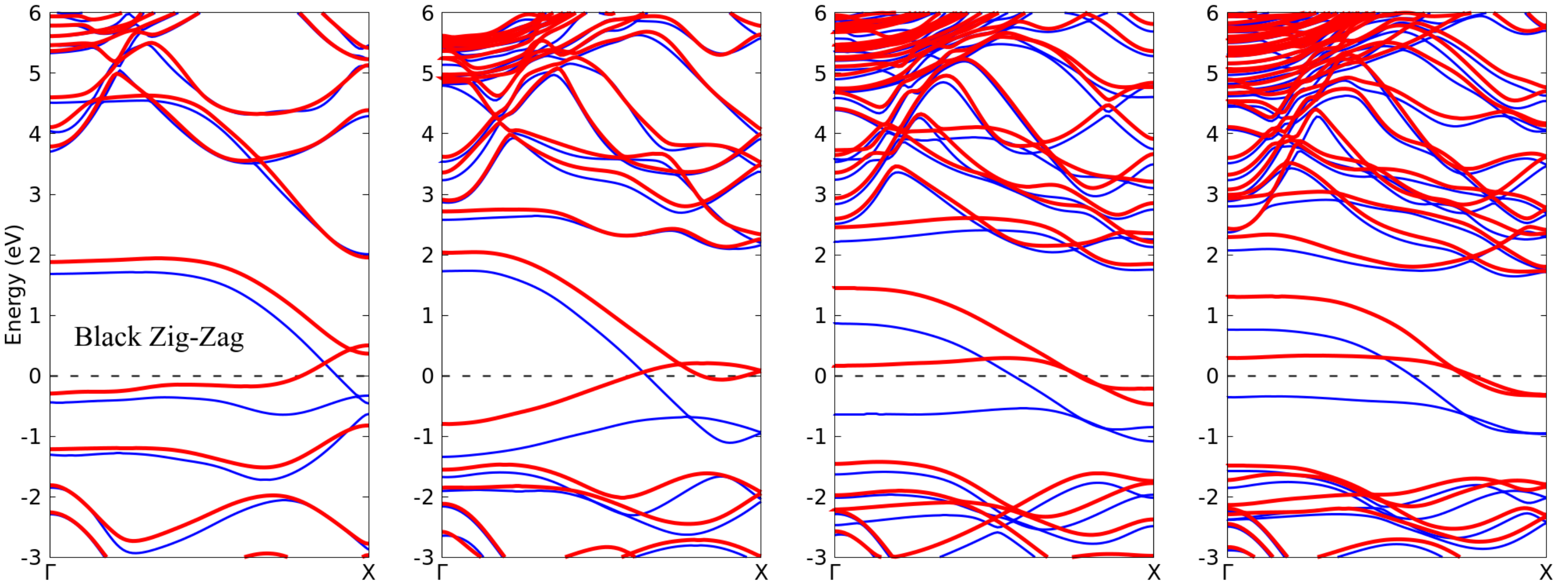}
    \caption{HSE06 band structure of black phosphorene nanoribbons. Upper panel:  armchair termination and widths of 
    a) \SI{2.5}{(L=2)}, b) \SI{5.2}{(L=3)}, c) \SI{6.7}{(L=4)} and d) \SI{8.2}(L=5){\AA}. Lower panel: zigzag termination and widths of  a) \SI{4.8}{(L=2)}, b) \SI{7.0}{(L=3)}, c) \SI{8.5}{(L=4)} and d) \SI{12.4}(L=5){\AA}. Blue (red) lines represent the up (down) spin channel. Fermi level is set at zero.}\label{fig:band_black_ribbons}
\end{figure}

Relaxed structures of black phosphorene nanoribbons are shown in Fig.\,\ref{fig:black_ribbons}(a) for  armchair termination and widths of \SI{2.5}{(L=2)}, \SI{5.2}{(L=3)}, \SI{6.7}{(L=4)} and \SI{8.2}(L=5){\AA} and  in Fig.\,\ref{fig:black_ribbons}(b) for zigzag termination and widths  of \SI{4.8}{(L=2)}, \SI{7.0}{(L=3)}, \SI{8.5}{(L=4)} and \SI{12.5}(L=5){\AA}.
The black nanoribbons with armchair termination and  L= 2, 3, 4, 5 show indirect band gap, seen in Fig.\,\ref{fig:band_black_ribbons} (upper panel). On the contrary, the zigzag edge-terminated ribbons contribute to metallic behavior for all widths. 
For smaller ribbons  with L= 2 and zigzag edges, a conductive behavior is found, as seen in the band structure of Fig.\,\ref{fig:band_black_ribbons} (lower panel). 

\begin{figure}[H]
    \centering
  {\includegraphics[width=0.9\linewidth]{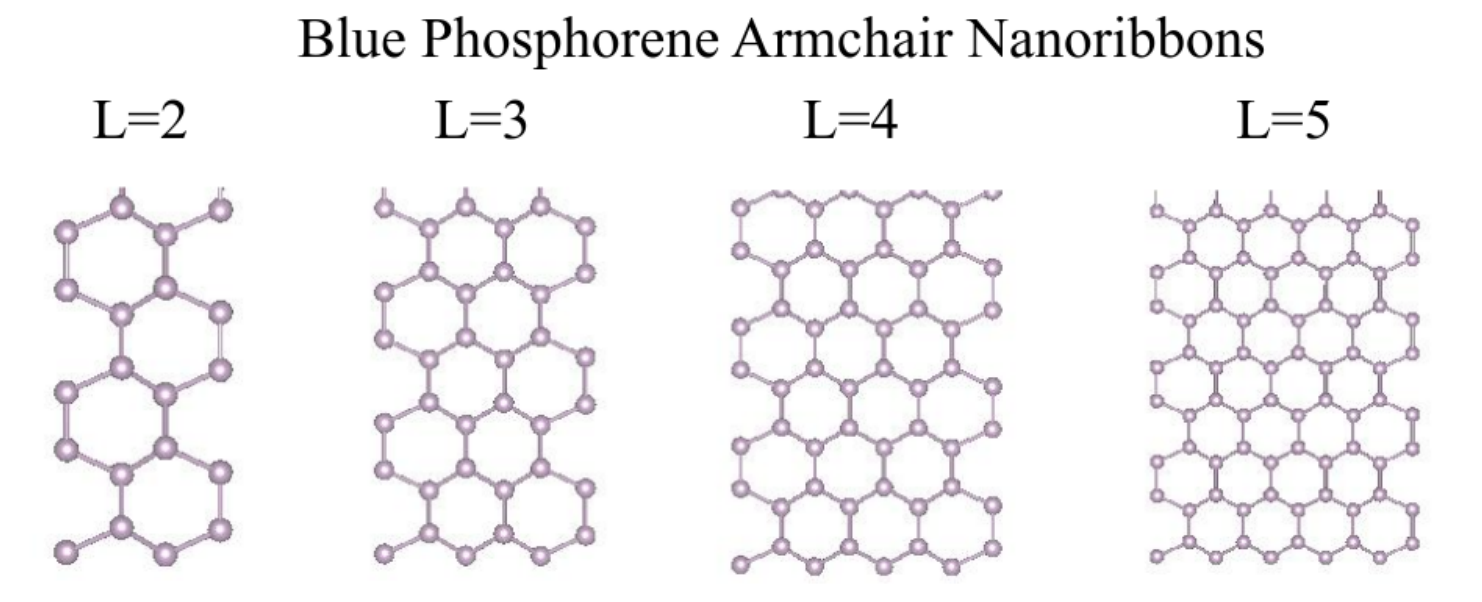}}\\
   {\includegraphics[width=0.9\linewidth]{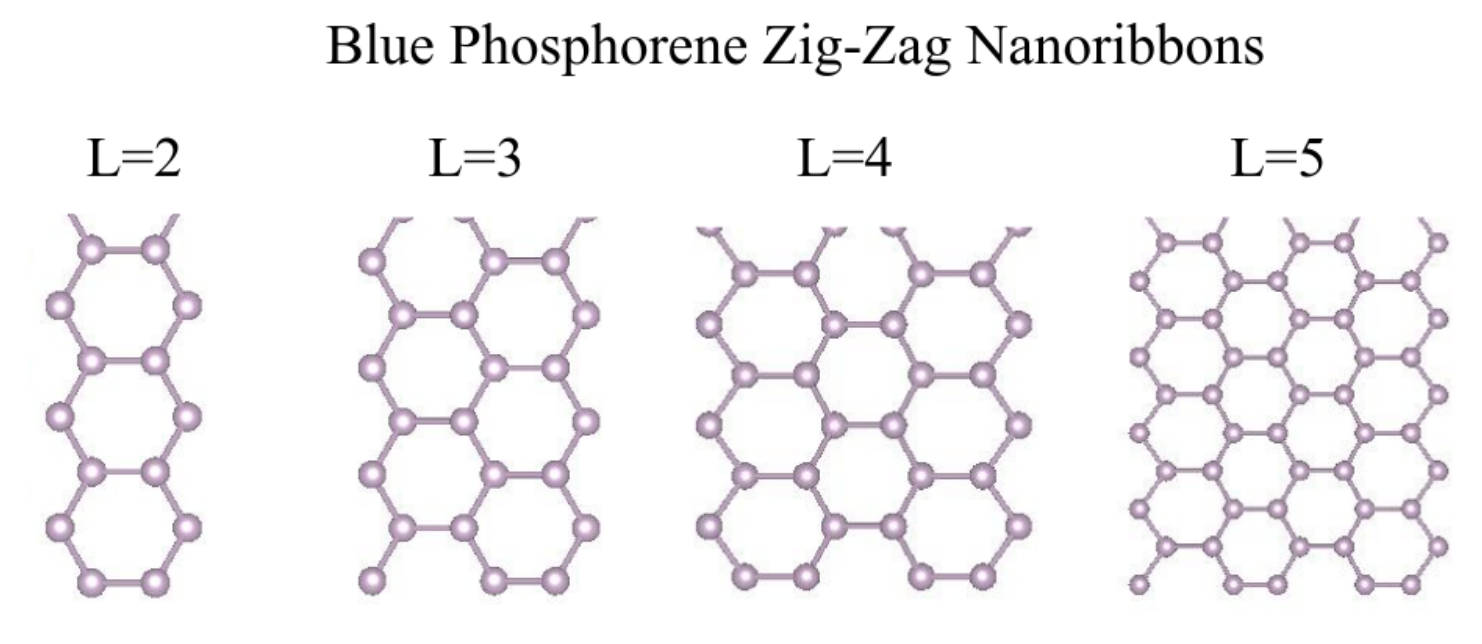}}
    \caption{\label{fig:blue_ribbons}Relaxed structure of blue phosphorene nanoribbons. a) growth along the zigzag direction with armchair edges and widths of \SI{5.4}({L=2)}, \SI{8.7}{(L=3)}, \SI{12}{(L=4)} and \SI{15.3}{(L=5)}\si{\angstrom}. b) growth along armchair direction with zigzag edges and widths of \SI{5.0}{(L=2)}, \SI{6.5}{(L=3)}, \SI{10.0}{(L=4)} and \SI{12.5}(L=5) \si{\angstrom}.}
\end{figure}

\begin{figure}[H]
    \centering
     \includegraphics[width=0.9\linewidth]{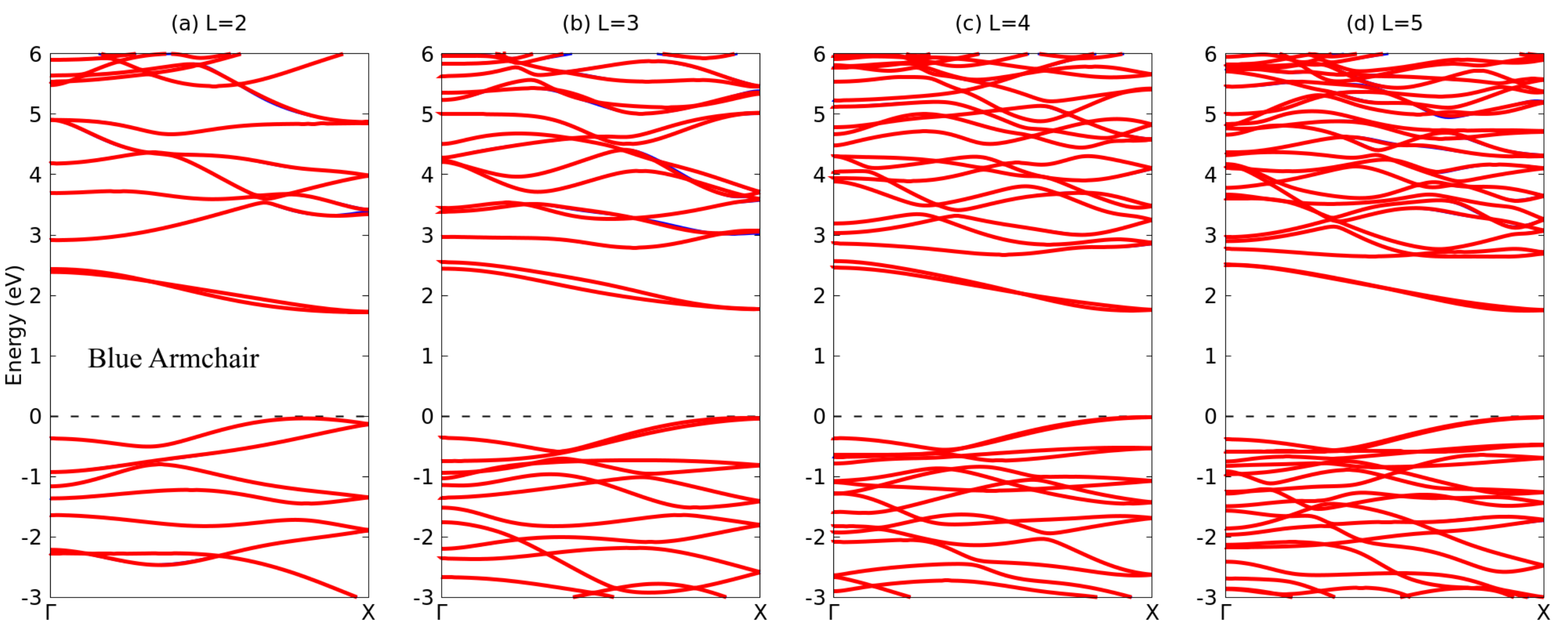}\\
    \includegraphics[width=0.9\linewidth]{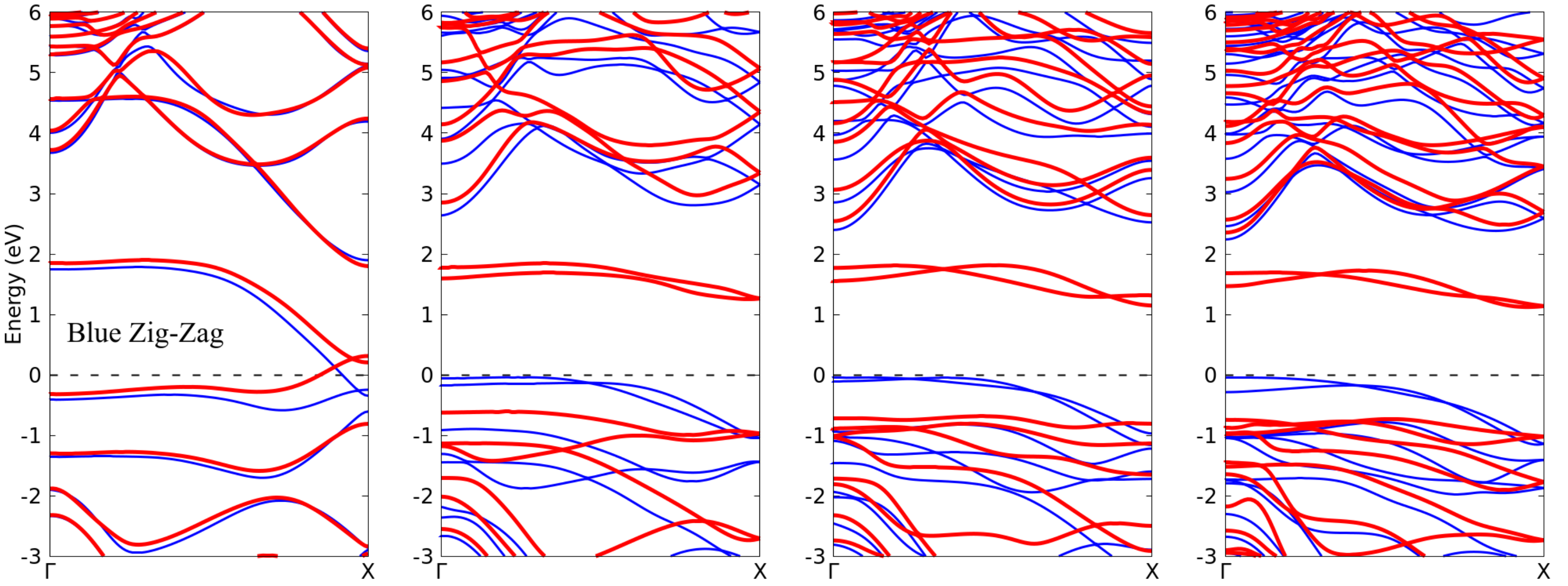}
    \caption{HSE06 band structure of blue phosphorene nanoribbons. Upper panel: growth along the zigzag direction with armchair edges and width of a) \SI{5.4}{(L=2)}, b) \SI{8.7}{(L=3)}, c) \SI{12}{(L=4)} and d) \SI{15.3}(L=5){\AA}. Lower panel: growth along the armchair direction with zigzag edges and widths of a) \SI{5.0}{(L=2)}, b) \SI{6.5}{(L=3)}, c) \SI{10.0}{(L=4)} and d) \SI{12.5}(L=5){\AA}. Blue (red) lines represent the up (down) spin channel. Fermi level is set at zero.}\label{fig:band_blue_ribbons}
\end{figure}

Relaxed structures of blue phosphorene nanoribbons are shown in Fig.\,\ref{fig:blue_ribbons}(a) for  armchair termination and widths of \SI{5.4}{(L=2)}, \SI{8.7}{(L=3)}, \SI{12}{(L=4)} and \SI{15.3}(L=5){\AA} and  in Fig.\,\ref{fig:blue_ribbons}(b) for zigzag termination and widths  of \SI{5.0}{(L=2)}, \SI{6.5}{(L=3)}, \SI{10.0}{(L=4)} and \SI{12.5}(L=5){\AA}.
As seen in Fig.\,\ref{fig:band_blue_ribbons} (upper panel) blue armchair ribbons with L = 2 have an indirect band gap, while for L = 3, 4 and 5 have direct gap at the X-point.   

Band structure of blue phosphorene ribbons with zigzag termination are shown in Fig.\,\ref{fig:band_blue_ribbons}(b). The smallest ribbon of width \SI{5.0}{\angstrom} (L=2) shown in Fig.\,\ref{fig:blue_ribbons} has a conductive character as seen in Fig.\,\ref{fig:band_blue_ribbons} (lower panel). As we increase the width L of the ribbons to \SI{6.5}{\angstrom} (L=3), \SI{10}{\angstrom} (L=4) and \SI{12.5}{\angstrom} (L=5) as seen in Fig.\,\ref{fig:band_blue_ribbons} we find indirect band gap semiconductors as seen in Fig.\,\ref{fig:band_blue_ribbons}.

Finally, we would like to mention that black phosphorus nanoribbons with armchair orientation have the lowest formation energy per atom, which is in agreement with the STM pictures reported in reference\,\cite{ribbon}. 

\section{Excitonic and optical Properties}

\subsection{Pristine phosphorene monolayers}

\begin{figure}[H]
    \centering
 \includegraphics[width=0.9\linewidth]{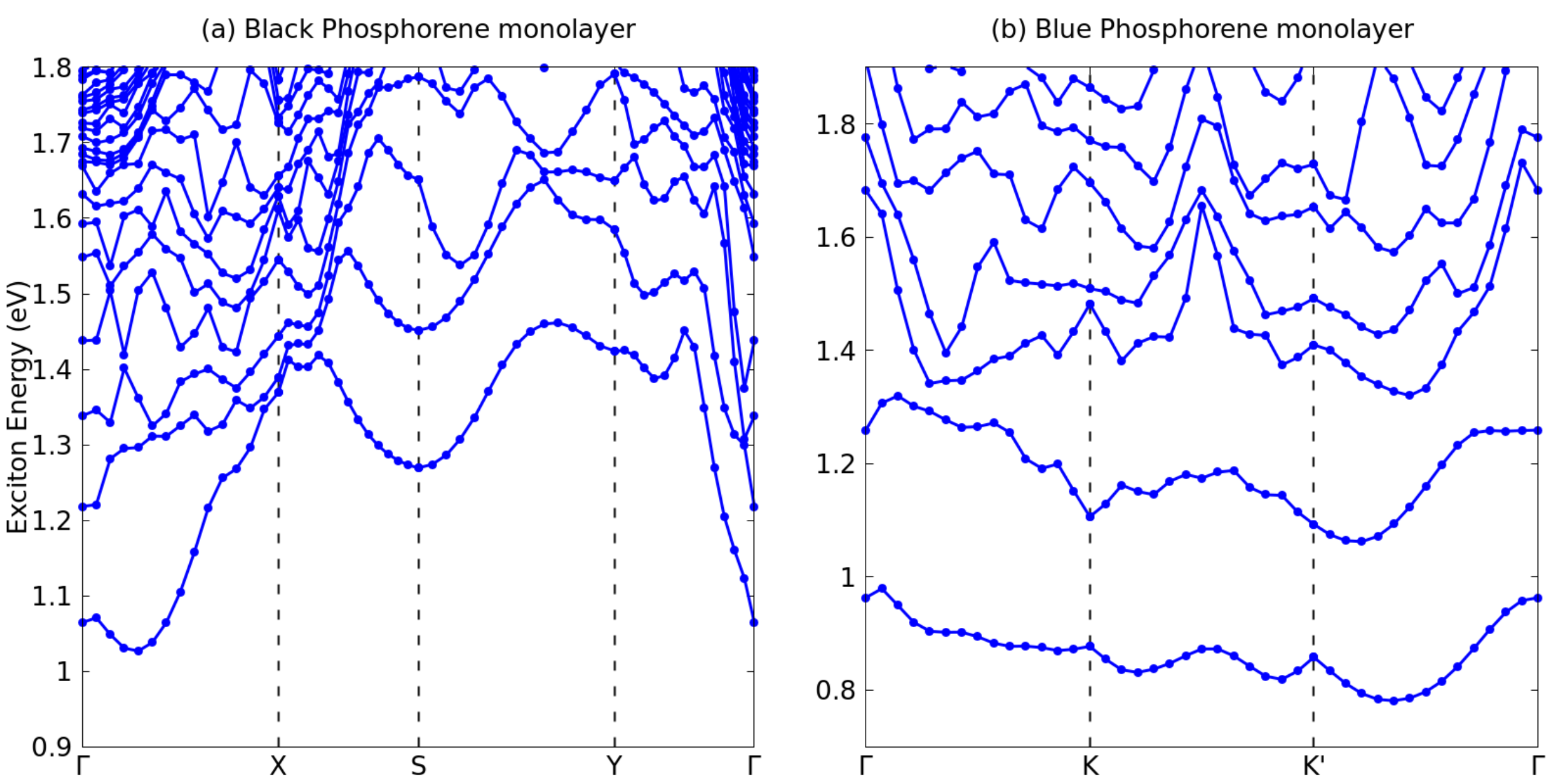}
    \caption{Exciton band structure of (a) black and (b) blue phosphorene monolayers}\label{fig:bands_bse_mono}
\end{figure}

From the exciton band structure shown in Fig.\,\ref{fig:bands_bse_mono} we can see that both monolayers have an indirect excitonic ground state located between $\Gamma$-S for black phosphorene with \SI{1.03}{\electronvolt} and between K'-$\Gamma$ points for blue phosphorene with \SI{0.78}{\electronvolt}, which results, when comparing with the fundamental band gap, in an exciton binding energy of \SI{564.18}{\milli\electronvolt}(\SI{1957.95}{\milli\electronvolt}) for the black(blue) phosphorene. The direct excitonic ground state for blue(black) phosphorene has excitation energies of \SI{1.06}{\electronvolt}(\SI{0.96}{\electronvolt}). Although blue and black phosphorene have the same chemical composition, we can see that the exciton binding energy in blue phosphorene is approximately four times higher than their black counterpart, with indicates that the Coulomb interaction is much stronger in the hexagonal buckled phase.

\begin{figure}[H]
    \centering
    \includegraphics[width=0.9\linewidth]{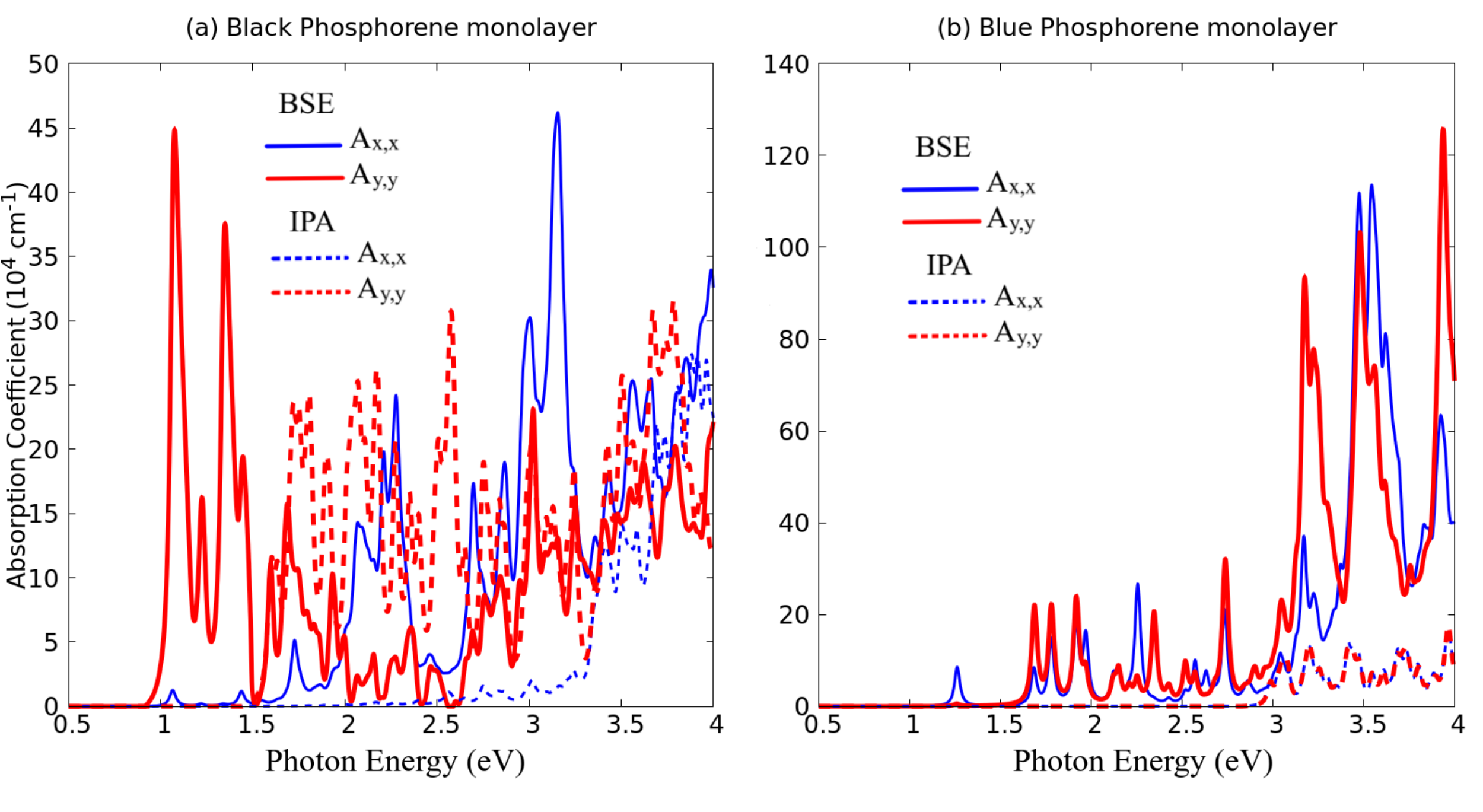}
    \caption{Absorption coefficient of (a) black and (b) blue phosphorene monolayers, at IPA (dashed lines) and BSE (solid lines) considering light polarization along $\hat{x}$ (blue curves) and $\hat{y}$ (red curves) directions.}
    \label{fig:bse_opt_mono}
\end{figure}

In Fig.\,\ref{fig:bse_opt_mono} we show the absorption coefficients for both black and blue phosphorene monolayers within BSE (with excitonic effects) and IPA (without excitonic effects) approximations. Comparing the dashed (IPA) and solid (BSE) curves, we find that the excitonic effects are significant for both phases. For the black phosphorene, we can see an optical anisotropy within both the IPA and BSE levels. This results in an optical band gap of \SI{2.27}{\electronvolt} for the polarization of $\hat{x}$ and \SI{1.59}{\electronvolt} for the polarization of $\hat{y}$ at the IPA level. Within BSE, despite the huge differences in the absorption spectrum for both directions of light polarization, the optical band gap is located at \SI{1.06}{\electronvolt}. Excitons in black phosphorene were found to be highly
anisotropic and strongly bound using polarization-resolved photoluminescence measurements at room temperature. Moreover, in this material, a quasiparticle bandgap of \SI{2.2}{\electronvolt} and an exciton binding energy of \SI{0.9}{\electronvolt}.\cite{NNano15} This corroborates with our results reported here, when the optical bandgap at IPA and BSE levels is compared, remembering that it is difficult to experimentally measure the indirect excitonic states.

On the other hand, for blue phosphorene we see that within IPA level, the system is optically isotropic, which means that regardless the light polarization, the optical response is the same. However, if excitonic effects are considered, we see significant changes in the absorption coefficient for peaks, which allows us to infer that the quasi-particle effects causes a small anisotropy in the optical response. The optical band gap is located in this case at \SI{2.97}{\electronvolt} within IPA and \SI{1.27}{\electronvolt} within BSE.  .

\subsection{Phosphorene nanoribbons}

\begin{figure}[H]
    \centering
    \includegraphics[width=0.9\linewidth]{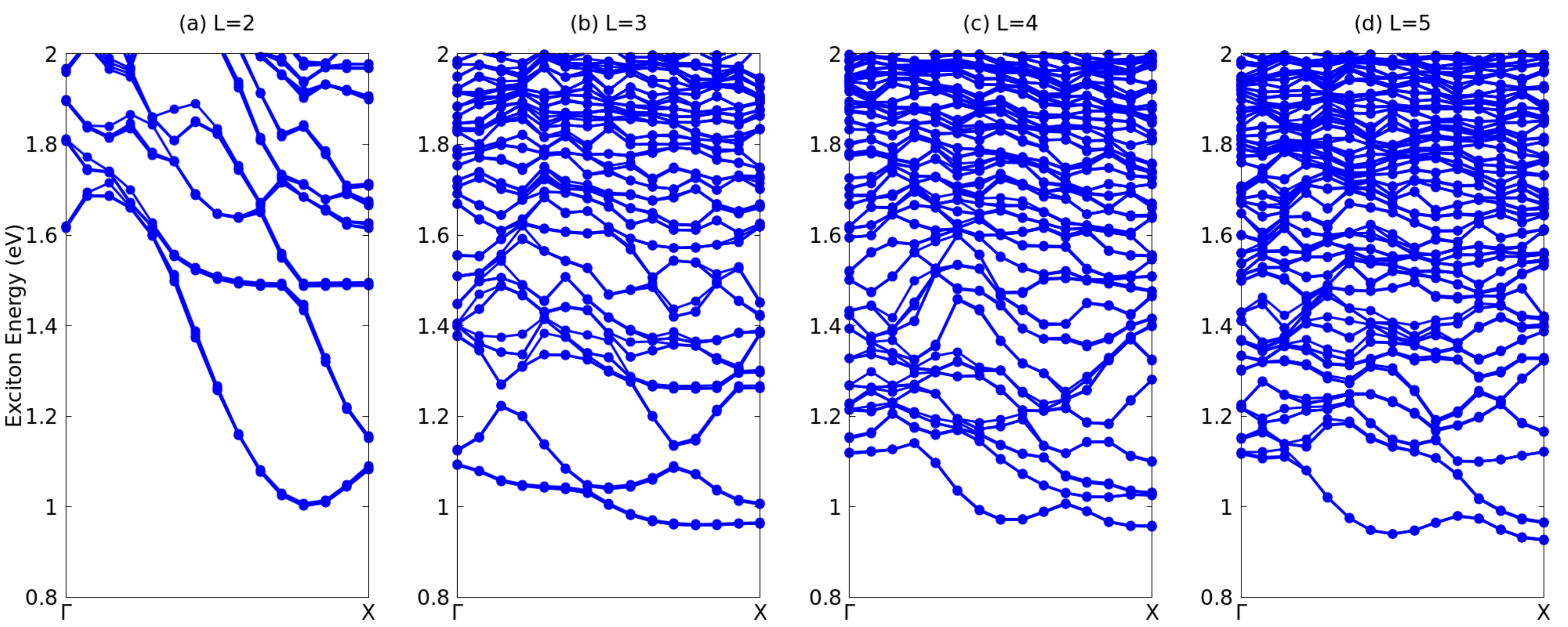}
    \includegraphics[width=0.9\linewidth]{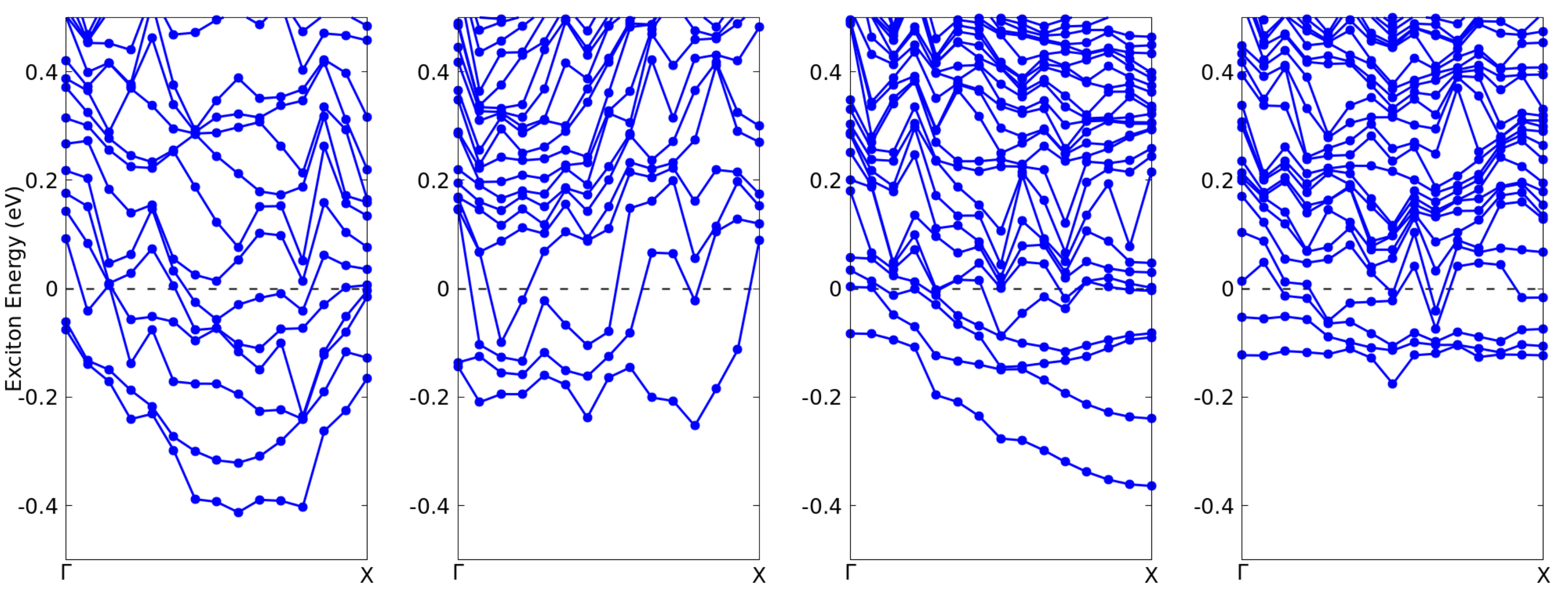}
    \caption{\label{fig:bands_bse_black_arm} Exciton band structure of black phosphorene nanoribbons. Upper panel: ribbons with armchair edges and width of a) \SI{2.5}{(L=2)}, b) \SI{5.2}{(L=3)}, c) \SI{6.7}{(L=4)} and d) \SI{8.2}{(L=5)}{\AA}. Lower panel: ribbons with zigzag edges and L of a) \SI{4.8}{(L=2)}, b) \SI{7.0}{(L=3)}, c) \SI{8.5}{(L=4)} and d) \SI{12.4}{(L=5)}{\AA}.}
\end{figure}

In Fig~\ref{fig:bands_bse_black_arm} we shown the excitonic band structure of black phosphorene armchair (upper panels) and zigzag (lower panels) nanoribbons. The armchair terminated nanoribbons has an indirect excitonic ground state. This means that the lowest excitonic state are formed for electrons and hole pairs at different \textbf{k}-points, the exciton binding energy, as shown in Table~\ref{tab:exciton_data}, decreases as the nanoribbon becomes wider. This is expected due to the reduction of the quantum confinement in the non-periodic direction. Similar  behaviour is found for the nanoribbons with zig zag edges. 

Despite the metallic behavior of the black zigzag nanoribbons, we obtained an exciton binding energy, higher than \SI{0}{\milli\electronvolt}. This means that the exciton binding energy is higher than the band gap, which suggests that these zig-zag nanoribbons becomes an excitonic insulator\,\cite{Kohn_462_1967,Jia_87_2022,Jiang_081408_2018}. Due to the spin composition of the lowest conduction and valence bands, our structures can be classified as a spin-triplet excitonic insulator.\cite{Jiang_166401_2020} This excitonic insulator behavior can explain why the exciton binding energy for black zigzag nanoribbons becomes much smaller than that for black phosphorous monolayers. As the quantum confinement is reduced, a possible semiconductor-excitonic insulator phase transition occurs.

\begin{figure}[H]
    \centering
    \includegraphics[width=0.9\linewidth]{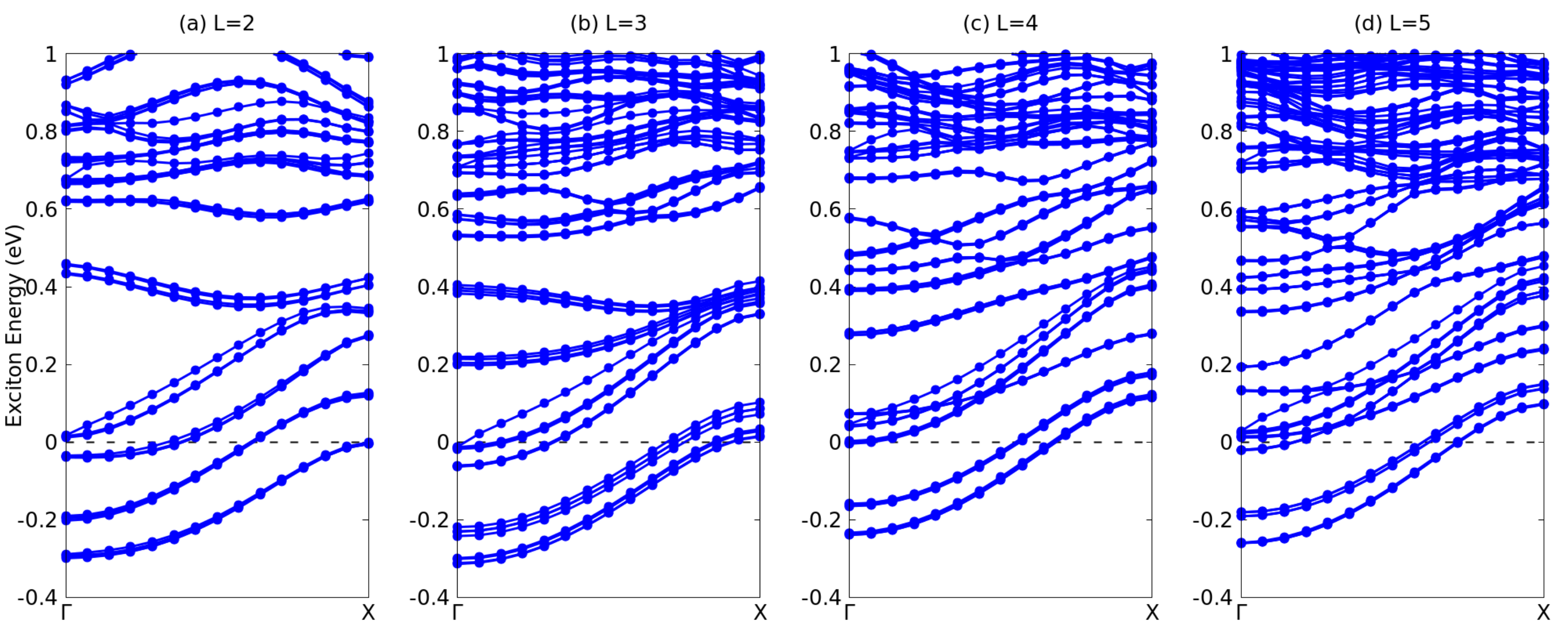}
\includegraphics[width=0.9\linewidth]{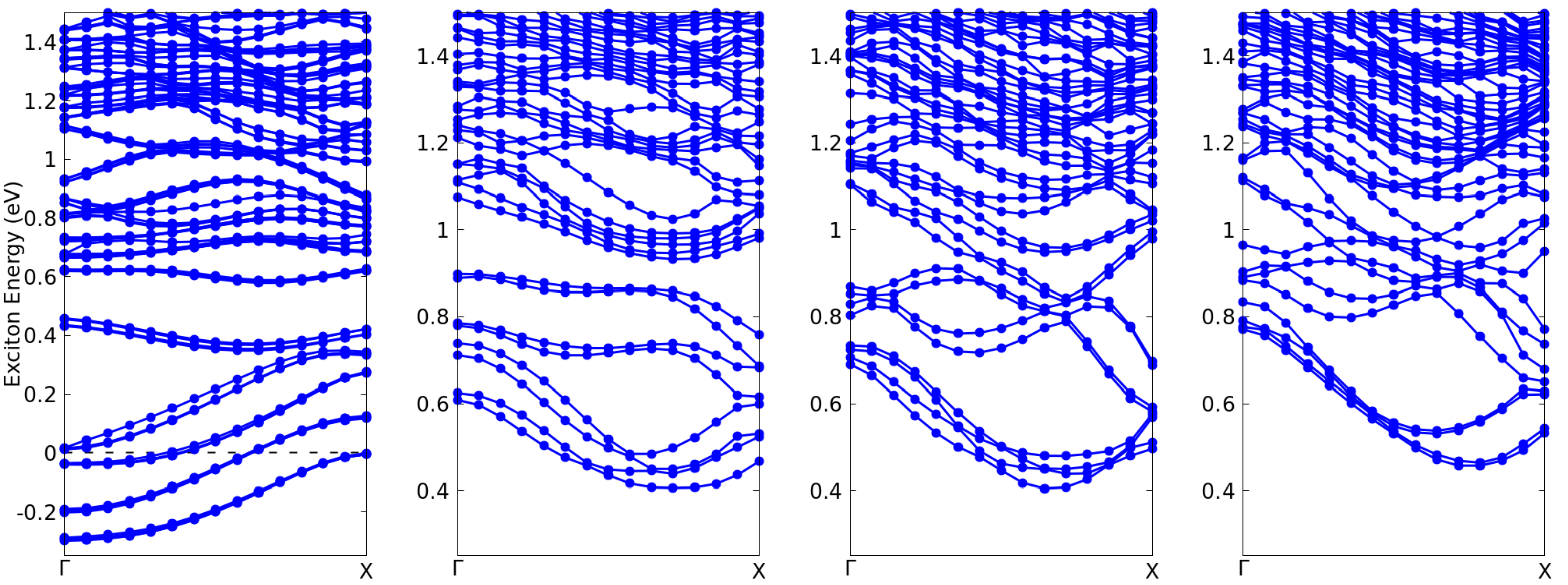}
    \caption{\label{fig:bands_bse_blue_arm}Exciton band structure of blue phosphorene. Uppper panel: ribbons with armchair edges and width of a) \SI{5.4}{(L=2)}, b) \SI{8.7}{(L=3)}, c) \SI{12.4}{(L=4)} and d) \SI{15.3}(L=5){\AA}. The same for ribbons with zigzag edges and width  of a) \SI{5.0}{(L=2)}, b) \SI{6.5}{(L=3)}, c) \SI{10.0}{(L=4)} and d) \SI{12.5}(L=5){\AA}.}
\end{figure}

The excitonic band structure for the blue phosphorene nanoribbons are shown in Fig~\ref{fig:bands_bse_blue_arm}, for the armchair (upper panels) and zig-zag (lower panels) terminations, the excitonic ground state indirect for all systems. From the armchair nanoribbons, regardless of the ribbon width, all armchair structures are excitonic insulators, with the same classification as the black phosphorene zigzag nanoribbons. Furthermore, an interest feature is found: the exciton binding energy $E_{\rm x}^{\rm b}$ shows negligible change with respect to the ribbon width, as reported in Table~\ref{tab:exciton_data}. This behaviour is the same shown for the electronic band gap of these systems, which suggests that the electron and hole pairs that compose these lower energy excitonic states are mostly edge states, resulting in high localized edge excitons. 

For blue phosphorene nanoribbons with zigzag edges, we find an indication of an excitonic insulator behavior for L = 2 and a trivial semiconductor behavior for larger widths. The excitonic insulator phase in L=2 occurs due the band gap closing as an effect of the quantum confinement.  Our results indicates a phase transition from excitonic insulator to semiconductor in a width between L=2 to L=3, where one expects a band gap opening. For L larger than 3, the exciton binding energy decreases as the quantum confinement decreases, as shown in Table~\ref{tab:exciton_data}. 

\begin{table}[H]
\caption{\label{tab:exciton_data} MLWF-TB+BSE electronic and excitonic properties for phosphorene nanoribbons: fundamental band gap $E_{\rm g}$, direct band gap, $E^{\rm d}_{\rm g}$, exciton ground state, $E_{\rm x}^{\rm gs}$, direct exciton ground state, $E_{\rm x}^{\rm gs}$ (d), and exciton binding energy, $E_{\rm x}^{\rm b}$, obtained from $E_{\rm g}-E_{\rm x}^{\rm gs}$. L is the ribbon width (see text).}
\centering 
\begin{tabular*}{12cm}{@{\extracolsep{\fill}}lccccccc}
    Structure & edges & L & $E_{\rm g}$ (\si{\electronvolt})& $E^{\rm d}_{\rm g}$ (\si{\electronvolt})& $E_{\rm x}^{\rm gs}$ (\si{\electronvolt})& $E{\rm x}^{\rm d}_{\rm gs}$ (\si{\electronvolt})&      $E_{\rm x}^{\rm b}$   (\si{\milli\electronvolt}) \\
    \hline
Black & armchair & 2   &1.68  &2.12  &1.00  &1.61  &677.83    \\
Black & armchair  & 3   &1.59  &1.62  &0.96  &1.09  &628.03    \\
Black & armchair  & 4   &1.48  &1.63  &0.96  &1.12  &525.71    \\
Black & armchair  & 5   &1.43  &1.60  &0.92  &1.12  &502.69    \\ \hline
Black & zigzag & 2   &0.00  &0.00  &\num{-0.41}  &\num{-0.08}  &412.63    \\
Black & zigzag & 3   &0.00  &0.00  &\num{-0.25}  &\num{-0.14}  &253.03    \\
Black & zigzag & 4   &0.00  &0.00  &\num{-0.36}  &\num{-0.08}  &363.97    \\
Black & zigzag & 5   &0.00  &0.00  &\num{-0.17}  &\num{-0.12}  &175.75    \\ \hline
Blue  & armchair  & 2   &1.76  &1.78  &\num{-0.30}  &\num{-0.30}  &2056.62    \\
Blue  & armchair  & 3   &1.80  &1.80  &\num{-0.31}  &\num{-0.31}  &2115.68    \\
Blue  & armchair  & 4   &1.76  &1.77  &\num{-0.24}  &\num{-0.24}  &2005.38    \\
Blue  & armchair  & 5   & 1.76 &1.77  &\num{-0.26}  &\num{-0.26}  &2026.57    \\ \hline
Blue  & zigzag & 2   &0.00  &0.00  &\num{-0.28}  &\num{-0.13}  &277.06    \\
Blue  & zigzag & 3   &1.29  &1.65  &0.40  &0.61  &885.35    \\
Blue  & zigzag & 4   &1.19  &1.60  &0.40  &0.69  &786.01  \\
Blue  & zigzag & 5   &1.17  &1.51  &0.46  &0.77  &709.43    \\ 
     \hline
    \end{tabular*}
\end{table}

In Table\,\ref{tab:exciton_data}  is possible to realize that the exciton binding energy does not vary significantly for blue nanoribbons with armchair edges. This might be an indication of strong localization of the electron-hole pair in the nanorribon edges. Recently, it it pointed out in Ref.\,\cite{Souvik2023} that black phosphorous edges atomic reconstructions can strongly confine excitons resulting in distinct emission features. The explanation is that local strain and screening can mediate such interactions, thus reducing the PL emission lines. Here we notice negligible bond length distortion between edge and bulk phosphorous atoms for L = 2-5 ribbons and therefore we can state that emission from edge states are not due to edge reconstruction.

\begin{figure}[H]
    \centering
    \includegraphics[width=0.9\linewidth]{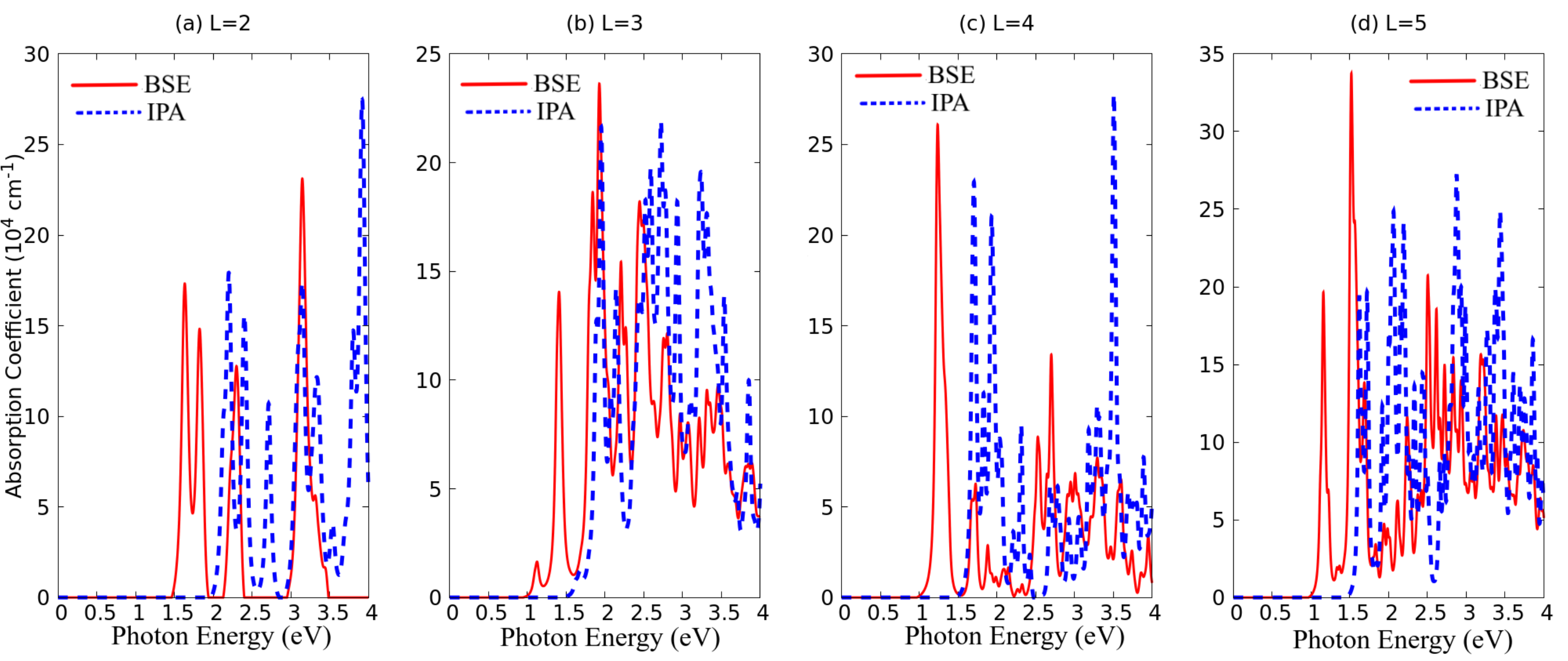}
     \includegraphics[width=0.9\linewidth]{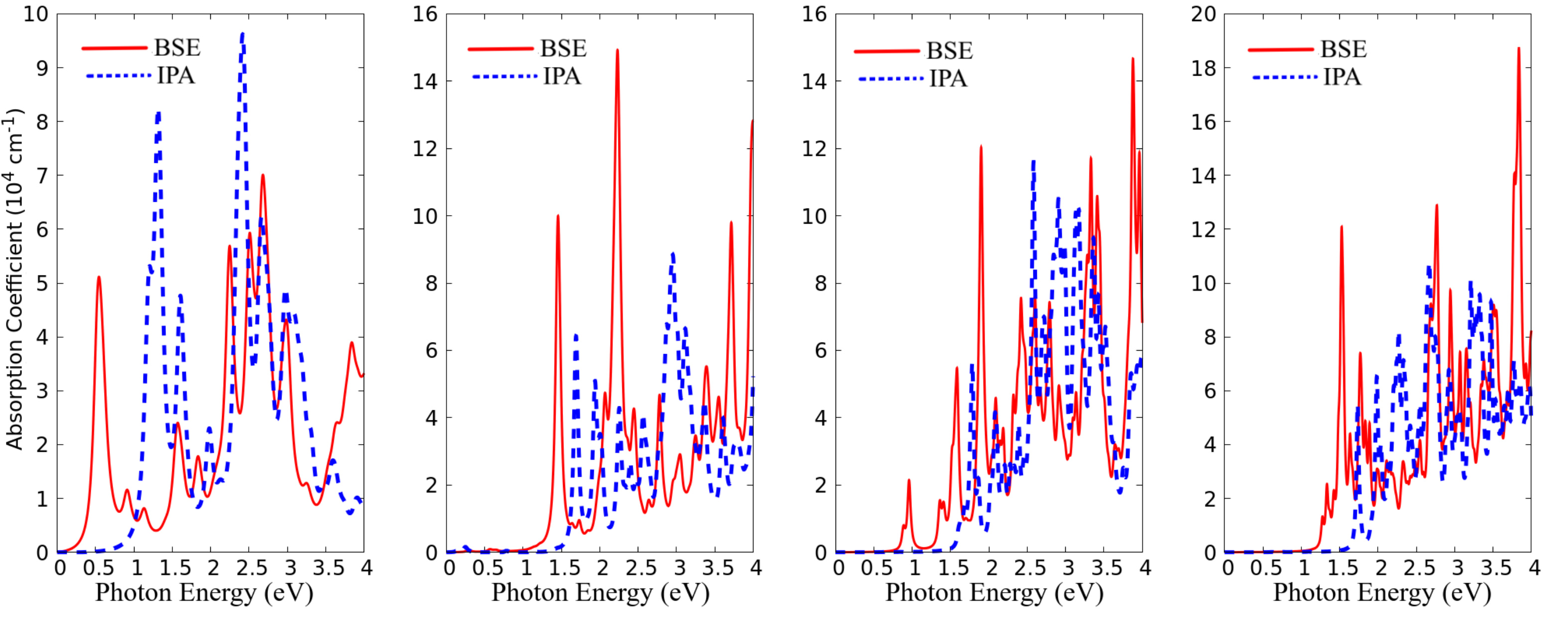}
    \caption{\label{fig:bse_opt_black_arm}Absorption coefficient for black phosphorene armchair nanoribbons calculated within IPA (blue dashed lines) and BSE (red solid lines). Upper panel: ribbon width of a) \SI{2.5}{(L=2)}, b) \SI{5.2}{(L=3)}, c) \SI{6.7}{(L=4)} and  d) \SI{8.2}{(L=5)}{\AA} (armchair edges). Lower panel: ribbon width of  a) \SI{4.8}{(L=2)}, b) \SI{7.0}{(L=3)}, c) \SI{8.5}{(L=4)} and d) \SI{12.4}{(L=5)}{\AA} (zigzag edges).}
\end{figure}

\begin{figure}[H]
    \centering
    \includegraphics[width=0.9\linewidth]{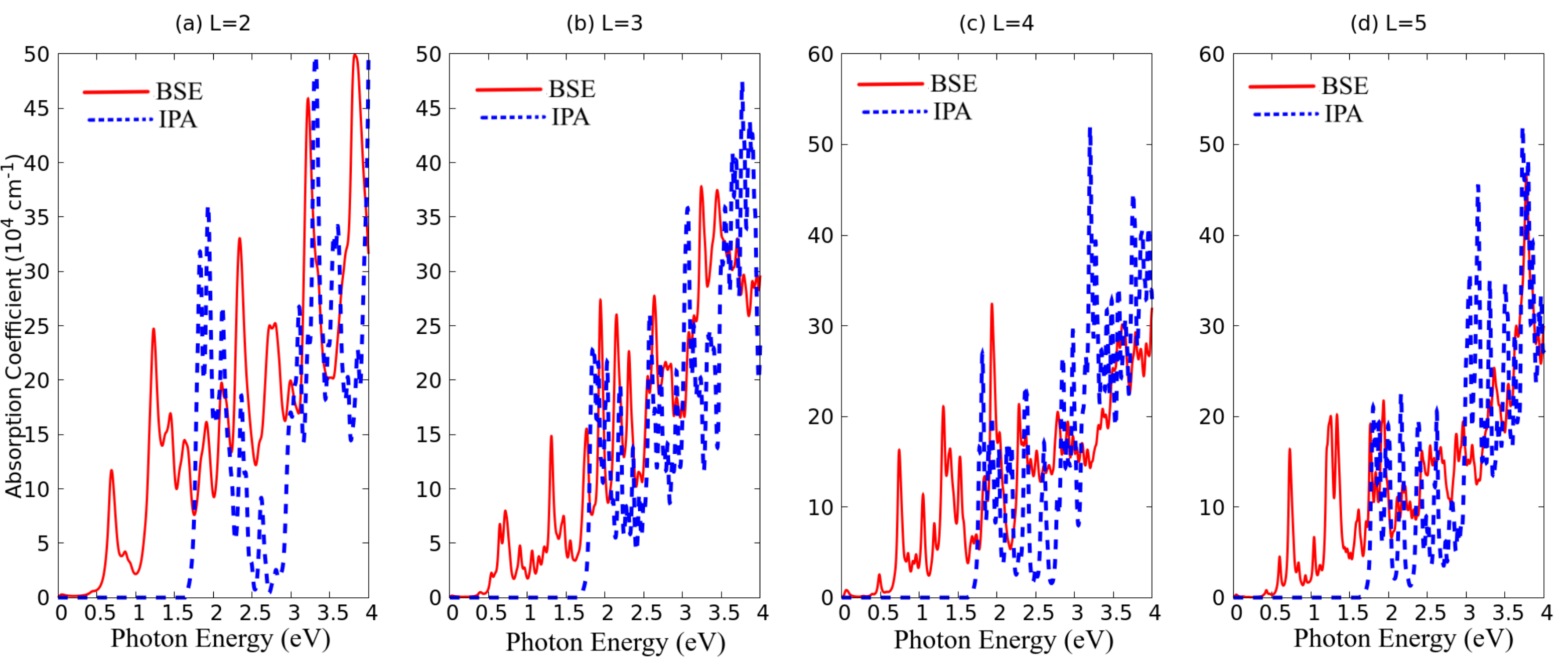}
    \includegraphics[width=0.9\linewidth]{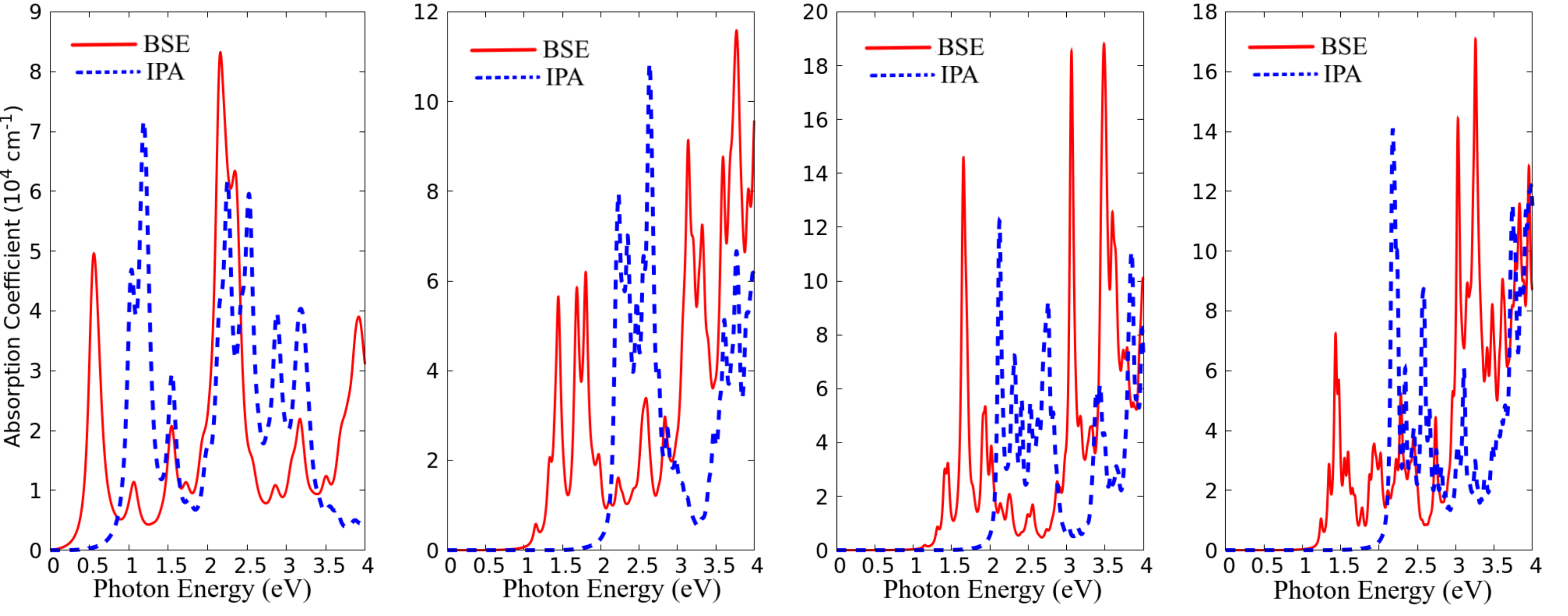}
    \caption{Absorption coefficient of the blue phosphorene armchair nanoribbons calculated within IPA (blue dashed lines) and BSE (red solid lines). Upper panel: Ribbon width of a) \SI{5.4}{(L=2)}, b) \SI{8.7}{(L=3)}, c) \SI{12.0}{(L=4)} and  d) \SI{15.3}(L=5){\AA} (armchair edges) and Lower panel: Ribbon width of a) \SI{5.0}{(L=2)}, b) \SI{6.5}{(L=3)}, c) \SI{10.0}{(L=4)} and d) \SI{12.5}(L=5){\AA} (zigzag edges).}\label{fig:bse_opt_blue_arm}
\end{figure}

The absorption coefficient of the investigated nanoribbons is shown in Figs.~\ref{fig:bse_opt_black_arm} and ~\ref{fig:bse_opt_blue_arm}, within both BSE and IPA levels of theory. Regardless of the nanoribbon we can observe that the quasi-particle effects are significant in the absorption spectrum shape and peaks, resulting in a red-shift. Interestingly enough, the excitonic insulators nanoribbons behave as semiconductor, having an optical band gap, as shown in Table~\ref{tab:opt_gap} higher than \SI{0}{\electronvolt}. The relation between optical band gap and the nanoribbon width are shown in Table~\ref{tab:opt_gap}. We notice that within IPA and BSE levels, the optical band gap has the lowest optical excitation. We consider the onset as the photon energy equals or higher than \SI{0}{\electronvolt} with an oscillator force higher than \SI{0.1}{\square\angstrom}.

\begin{table}[H]
\centering
\caption{Optical band gap within IPA ($E_{\rm gopt}^{\rm IPA}$) and BSE ($E_{\rm gopt}^{\rm BSE}$) levels for phosphorene nanoribbons. The optical band gap is defined as the first optical transition, with photon energy higher or equals to \SI{0}{\electronvolt}. The oscillator force is set to be higher than \SI{0.1}{\square\angstrom}. L is the ribbon width (see text).}
    \setlength{\tabcolsep}{5pt}
\begin{tabular*}{12cm}{@{\extracolsep{\fill}}lcccc}
    \toprule
   structure & edge & L  & $E_{\rm gopt}^{\rm IPA}$ (\si{\electronvolt})& $E_{\rm gopt}^{\rm BSE}$ (\si{\electronvolt})\\ \midrule
Black & armchair & 2   &2.12  &1.61   \\
Black &  armchair &3   &1.62  &1.09    \\
Black & armchair & 4   &1.63  &1.21    \\
Black & armchair & 5   &1.61  &1.11    \\ \hline
Black &  zigzag  & 2   &1.14  &0.50   \\
Black &  zigzag & 3   & 0.15  &0.28    \\
Black &  zigzag & 4   &1.60  &0.88   \\
Black &  zigzag & 5  &1.73  &1.27   \\ \hline
Blue  & armchair & 2   &1.78  &0.01   \\
Blue & armchair & 3  &1.80  &0.38    \\
Blue & armchair & 4   &1.76  &0.03   \\
Blue & armchair & 5   &1.77  &0.02   \\ \hline
Blue & zigzag & 2  &1.01  &0.52    \\
Blue & zigzag & 3   &2.19  &1.15    \\
Blue & zigzag & 4   &2.11  &1.15  \\
Blue & zigzag & 5   &2.16  &2.10    \\ 
     \bottomrule
    \end{tabular*}
    \label{tab:opt_gap}
\end{table}

\section{Conclusions}

Our electronic structure calculations shows that both monolayered phases are semiconductors. For the nanoribbons we see that black phosphorene zig-zag nanoribbons are metallic. Similar bheavior is found for thin  blue phosphorene zig-zag and armchair nanoribbon also metallic.  For the semiconductor nanoribbons the electronic band gap decreases as the nanoribbon width increases.

The solution of the Bethe-Salpeter equation shows that blue phosphorene monolayer has an exciton binding energy four times higher than the black phosporene  counterpart. Furthermore, both monolayers shows a different linear optical response due light polarization, being black phosphorene highly anisotropic. We find a similar, but less pronounced, optical anisotropy is also seen in blue phosphorene monolayer, caused exclusively by the quasi-particle effects.

As a general behavior, the exciton binding energy decreases as the ribbon width increases, which highlights the importance of quantum confinement effects.  In addition to large band gap found  for some of the investigated structures, blue nanoribbons with zig-zag and armchair edges shows a spin-triplet excitonic insulator behavior. 
  Our calculations reveal exciting features in these nanomaterials and therefore provides important advances in the understanding of quasi-one- dimensional black phosphorus.

\section{Acknowledgements}

We acknowledge the financial support from the Brazilian funding agency CNPq under grant numbers 313081/2017-4, 305335/2020-0, 309599/2021-0 and 408144/2022-0. We thank computational resources from LaMCAD/UFG, Santos Dumont/LNCC, CENAPAD-SP/Unicamp (project number 897) and Lobo Carneiro HPC (project number 133). A.C.D also acknowledge FAPDF grants numbers 00193-00001817/2023-43 and 00193-00002073/2023-84. 

\section{Electronic Support Information}
The atomic structures data, complementary data for the figures shown in manuscript and additional technical details are reported within the Electronic Support Information (SI).

\section{Conflict of interest statement}
The authors declare no conflict of interest.

\section{Data availability statement}
The atomic structures data, complementary data for the figures shown in manuscript and additional technical details are reported within the Electronic Support Information (SI).

\section{Ethics statement}

It does not apply.

\bibliography{refs.bib}

\end{document}